\documentclass[fleqn,usenatbib]{mnras}

\usepackage{newtxtext,newtxmath}
\usepackage{bm} 

\usepackage[T1]{fontenc}

\usepackage{graphicx}	
\usepackage{amsmath}	
\usepackage{multirow}   

\defcitealias{Smith2016merger}{S16}

\title[Imprint of mergers on satellite planes]{The imprint of galaxy mergers on satellite planes in a cosmological context}

\author[K. J. Kanehisa et al.]{
Kosuke Jamie Kanehisa$^{1,2,3}$\thanks{E-mail: kkanehisa@aip.de},
Marcel S. Pawlowski$^{1}$,
Oliver M\"{u}ller$^{4}$
\\
$^{1}$Leibniz-Institut für Astrophysik Potsdam (AIP), An der Sternwarte 16, 14482 Potsdam, Germany\\
$^{2}$Institut für Physik und Astronomie, Universität Potsdam, Karl-Liebknecht-Straße 24/25, D-14476 Potsdam, Germany\\
$^{3}$Department of Physics, University of Surrey, Guildford GU2 7XH, UK\\
$^{4}$Institute of Physics, Laboratory of Astrophysics, Ecole Polytechnique Fédérale de Lausanne (EPFL), 1290 Sauverny, Switzerland\\
}

\date{Accepted 2023 June 15. Received 2023 May 20; in original form 2022 December 20}

\pubyear{2023}

\begin{document}
\label{firstpage}
\pagerange{\pageref{firstpage}--\pageref{lastpage}}
\maketitle

\begin{abstract}
Flattened and kinematically correlated planes of dwarf satellite galaxies have been observed in the Local Volume.
The slinging out of satellites during host galaxy mergers has been suggested as a formation mechanism for these peculiar structures. 
We statistically examined the impact of major mergers on present-time satellite systems for the first time in a full cosmological context using the IllustrisTNG suite of hydrodynamic simulations.
Mergers with mass ratios above $1/3$ generally have a negligible or adverse impact on the phase-space correlation of observationally motivated satellites.
Even high-angular momentum mergers are inefficient at slinging satellites outward due to the extended nature of simulated satellite distributions.
Furthermore, any potential merger imprint is partially washed out by post-merger accretion of satellites, while satellites bound to the merging haloes since the merger's beginning are disrupted and stripped of mass -- minimising the merger's influence on the present-time distribution of the most massive satellites after $2-5\,\mathrm{Gyr}$.
Constraining our sample to satellites bound to their host throughout the full duration of their system's last merger, we recover no particular improvement in their phase-space correlation.
Instead, such participant satellites experience a contraction of their radial distribution during and after the merger, resulting in smaller absolute plane heights (but comparable axis ratios). 
Overall, major mergers do not appear to form correlated planes in a statistical sample.
Mergers that efficiently transfer their angular momentum to satellite distributions can marginally enhance their phase-space correlation, but cannot form highly flattened and orbitally coherent configurations as observed in our local Universe.
\end{abstract}




\section{Introduction}
\label{sec:s1}

It has been known since \citet{Kunkel1976mw} and \citet{Lynden-Bell1976mw} that the satellite galaxies of the Milky Way align along an extended planar distribution, now referred to as the Vast Polar Structure (VPOS). While generally dominated by the 11 so-called classical satellites \citep{Kroupa2005dos, Metz2007distribution}, this plane-of-satellites is further traced by many of the Milky Way's fainter dwarfs \citep{Metz2009newsats, Pawlowski2015newsats}.
Furthermore, proper motions measured for many of the VPOS's constituents point towards a common co-rotational motion aligned with the plane \citep{Metz2008pms, Pawlowski2013pms, Fritz2018dr2, Pawlowski2020dr2, Li2021dr3}, and may reflect an inherent rotational stability.

A similarly anisotropic structure, now called the Great Plane of Andromeda (GPoA), has also been discovered around M31 \citep{Koch2006m31, McConnachie2006m31, Metz2007distribution}. Here, 15 out of 27 satellites form a highly flattened subsample with a plane height as low as 13 kpc \citep{Ibata2013m31, Conn2013m31}.
Owing to the GPoA's nearly edge-on orientation, the on-plane satellites' radial velocities hint at a strong degree of co-rotation (though a correlation in radial velocities does not necessarily confirm a kinematically coherent plane, see \citealt{Gillet2015gpoa} and \citealt{Buck2016gpoa}). Recent proper motions for two of M31's dwarfs lends credence to a rotationally supported GPoA \citep{Sohn2020m31}, although the long-term stability of the plane is uncertain \citep{Fernando2017gpoa, Fernando2018gpoa}.

Beyond the Local Group, Centaurus A has been found to also host a plane-of-satellites \citep{Tully2015cena, Muller2016cena}, which is oriented nearly edge-on to the Milky Way and may be rotationally supported \citep{Muller2021cena}. The plane is notable not solely due to its thickness, which at around 130 kpc \citep{Muller2019casp} is significantly wider than the VPOS and GPoA, but in conjunction with a strong implied degree of co-rotation from the system's radial velocities \citep{Muller2018casp, Muller2021casp}.
Many other systems in the Local Volume ($D < 10\,\mathrm{Mpc}$) also demonstrate indications of hosting a flattened satellite structure (M81: \citealt{Chiboucas2013m81}, M101: \citealt{Muller2017m101}, M83: \citealt{Muller2018m83}, NGC253: \citealt{Martinez-Delgado2021ngc253}). Such anisotropic structures may be ubiquitous in the local Universe (\citealt{Ibata2014sdss, Ibata2015sdss, Heesters2021matlas, Paudel2021ngc2750}, though see \citealt{Phillips2015sdss, Cautun2015sdss}).

The strong anisotropy inherent in local satellite systems has been argued to be inconsistent with structure formation in concordance cosmology, wherein the hierarchical clustering of dark haloes produce near-isotropic satellite distributions \citep{Kroupa2005dos}. Preferential directions of infall arising from the orientation of nearby filaments, the accretion of satellites as gravitationally bound groups, and the inherent triaxiality of CDM haloes can all introduce a notable degree of anisotropy \citep{Zentner2005cdm, Libeskind2005cdm}.
Regardless, satellite distributions as flattened and kinematically correlated as the observed Local Volume planes are rare ($<0.2$ per cent) in CDM simulations \citep[e.g.][]{Pawlowski2014cdm, Pawlowski2020dr2, Ibata2014cdm, Muller2021casp}, although the degree of tension this invokes is still a topic of debate \citep[e.g.][]{Bahl2014, Cautun2015elsewhere, Samuel2021fire}.

\begin{figure*}
	\includegraphics[width=0.322\textwidth]{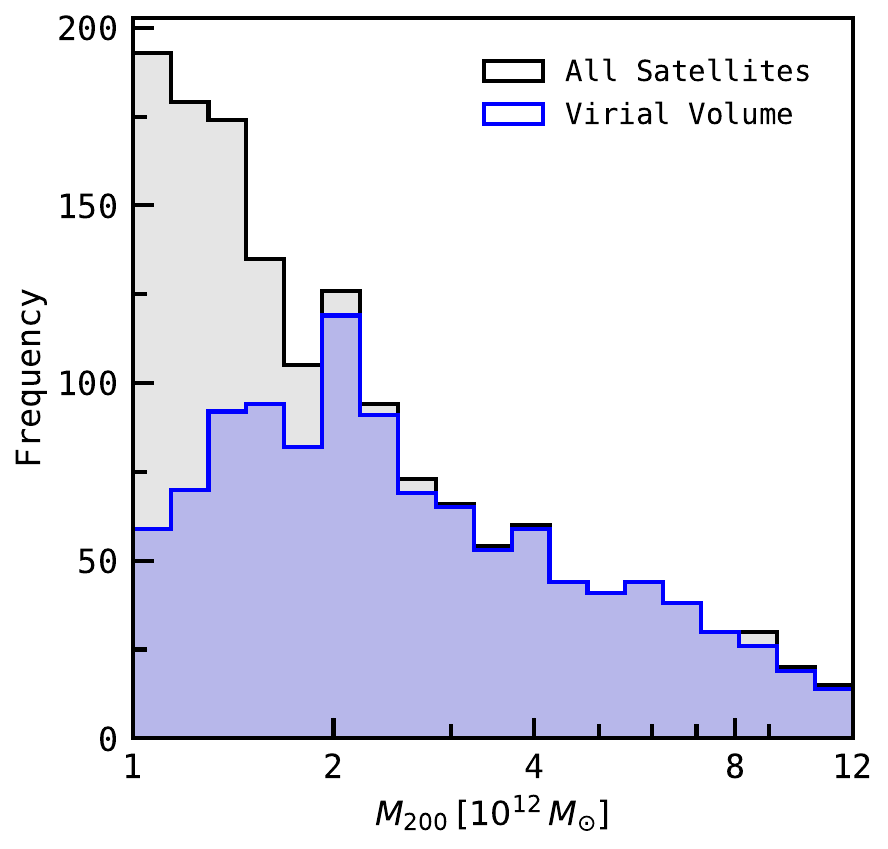}
	\includegraphics[width=0.32\textwidth]{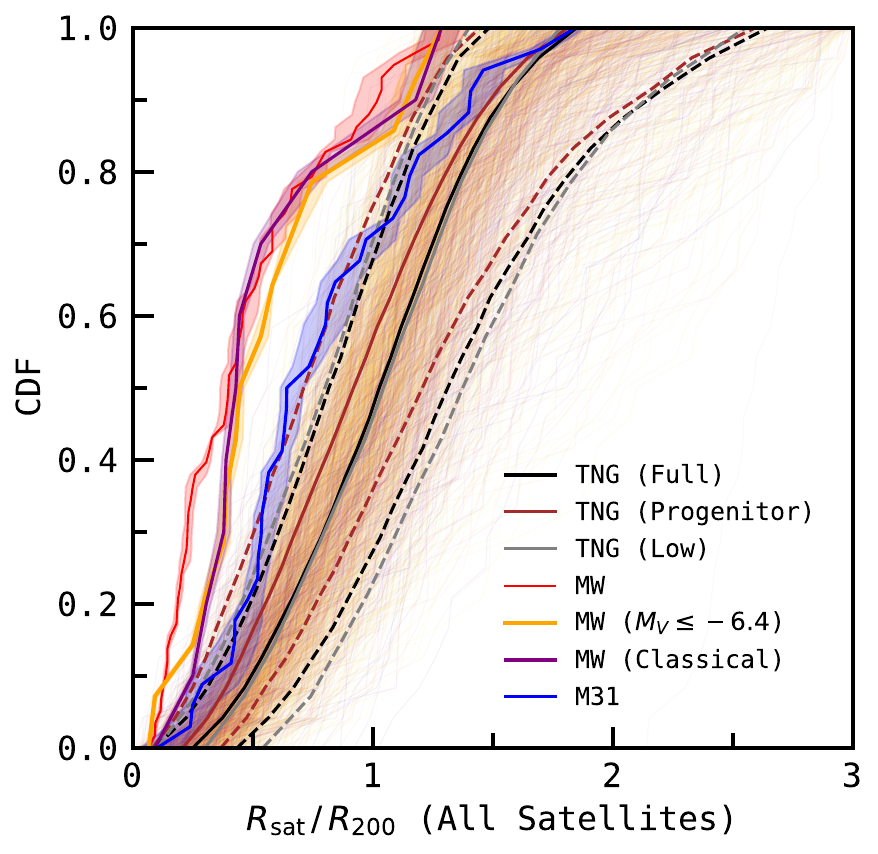}
	\includegraphics[width=0.327\textwidth]{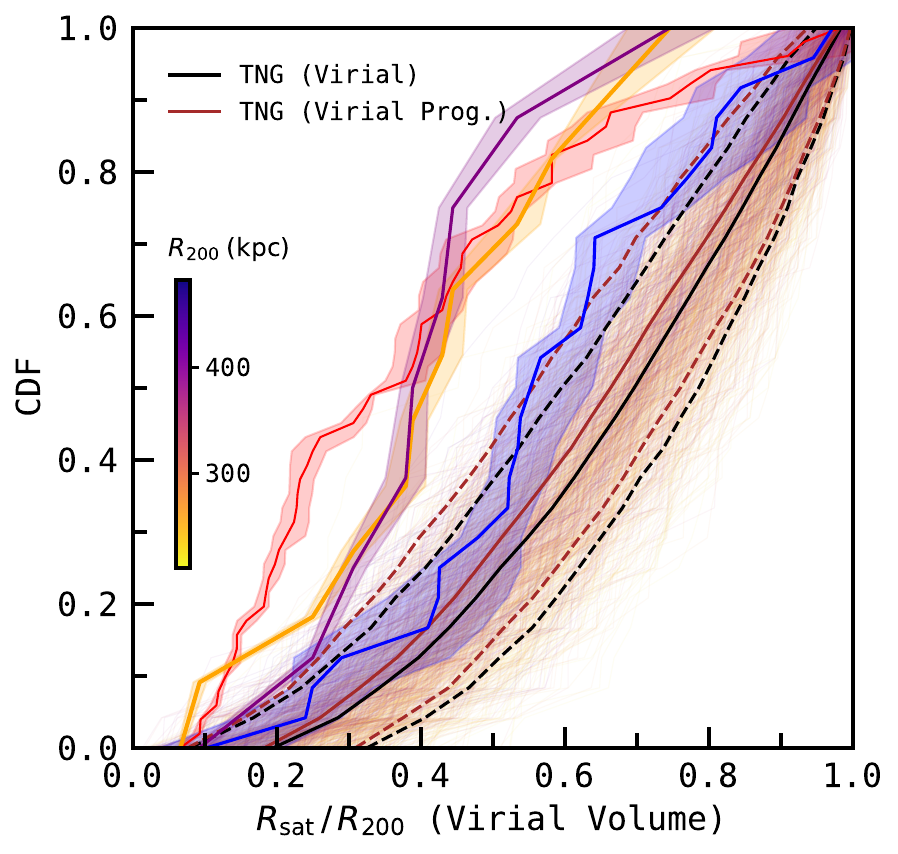}
    \caption{
        Present-time properties of 1571 host galaxies and their satellites sampled from TNG100-1. 
        \textbf{Left Panel:} Distribution of host halo masses. The grey histogram represents our halo sample when selecting the 25 most massive bound subhaloes, while the blue histogram represents that when only sampling subhaloes within their host halo's virial radius. Many low-mass hosts ($M_{200}<2\times10^{12}\,M_{\odot}$) are excluded due to an insufficient satellite population when restricting our search to the virial volume.
        \textbf{Central Panel:} Satellite radial distributions normalised to their host halo's virial radius. Each thin line represents a single simulated system, and is coloured with respect to their halo radius. The solid black line represents their median distribution, while 10th and 90th percentiles are indicated by the two dashed black lines. Brown and grey lines in the same manner represent simulated results for the \emph{progenitor} and \emph{low} samples respectively (see Table~\ref{tab:s2_samplesize} for sample definitions). Red and blue lines represent the observed radial distribution of the Milky Way \citep{McConnachie2012lg} and M31 \citep{Savino2022m31} satellite systems respectively, while areas shaded in their corresponding colours account for distance uncertainties. In addition, the orange line represents the MW satellites with an absolute magnitude cut ($M_V \leq -6.4$) motivated by the M31 system's lowest-luminosity satellite, while the purple line shows the distribution of the 11 classical MW satellites.
        \textbf{Right Panel:} Satellite radial distributions when restricting the search area to the virial volume.
    }
    \label{fig:s2_sample}
\end{figure*}

\begin{table*}
	\centering
	\caption{Sizes of four samples of simulated host-satellite systems used in this work. Each sample is given a designation for reference and a corresponding selection criteria. The remaining columns display each sample's total number of systems, \emph{merger-type} systems that have experienced a major merger within the last 5 Gyr (see Section~\ref{sec:s3}), systems with sufficient participant satellites that were bound to progenitors of their current host halo before the last merger began (see Section~\ref{sec:s4_participants}), and participant systems additionally drawn from TNG50.}
	\vspace*{-1mm}
	\begin{tabular}{llllll}
		\hline
		Sample & Criteria & $N_{\mathrm{total}}$ & $N_{\mathrm{merger}}$ & $N_{\mathrm{part}}$ & $N_{\mathrm{part (TNG50)}}$\\
		\hline
		full & $M_{200}=[1,12]\times10^{12}M_{\odot}$, no companion with $>0.25M_{200}$ within $3R_{200}$, 25 satellites & 1521 & 360 & 25 & 16\\
		isotropic & Isotropically drawn positions and velocity vectors with magnitudes from \emph{full} sample & 1521 & -- & -- & --\\
		virial & Same as the \emph{full} sample, but restrict satellite search radius to $R_{200}$ & 1109 & 273 & 22 & 7\\
		progenitor & Same as the \emph{full} sample, but rank satellites by their maximum past progenitor $M_{\mathrm{dm}}$ & 1513 & 358 & 24 & 16\\
		virial progenitor & Adopt both $R_{200}$ search radius and maximum past progenitor $M_{\mathrm{dm}}$ ranking & 1062 & 262 & 21 & 7\\
		low & Same as the \emph{full} sample, but only consider the 15 most massive satellites & 1574 & 371 & 53 & 21\\
		\hline
	\end{tabular}
	\label{tab:s2_samplesize}
\end{table*}

At present, there is no clear consensus on the mechanism through which such correlated planes are formed.
The anisotropic accretion of satellites through nearby filaments \citep{Lovell2011filament, Libeskind2011filament} was raised to explain the phase-space correlation of observed satellite distributions, but filaments feeding Milky Way-mass haloes are wider than the hosts' virial radii and cannot produce thin planes with heights of $10-30\,\mathrm{kpc}$ \citep{Pawlowski2013lgplanes, Pawlowski2015newsats}.
The infall of bound galaxy groups produce subsets of satellites sharing similar orbital properties \citep{Lynden-Bell1995group, DOnghia2008group, Angus2016group}, but such groups are extended \citep{Metz2009group} and consist of only $2-5$ galaxies \citep{Li2008group, Wetzel2015group, Shao2018group} -- too few to reproduce the rich membership of the Local Group planes. In addition, both effects are self-consistently included in CDM simulations, wherein correlated planes-of-satellites remain rare.
Alternatively, it is conceivable that the observed LG satellites are devoid of dark matter entirely, and are instead tidal dwarfs formed from the debris of a past galaxy interaction \citep{Metz2007tdg, Pawlowski2012vpos, Hammer2010tdg, Hammer2013tdg}. Nevertheless, this scenario would represent a severe deficiency of primordial, dark matter-dominated dwarfs as expected by the Dual Dwarf Galaxy Theorem \citep{Kroupa2012theorem, Dabringhausen2013tdg}, or else require the invocation of alternate cosmologies such as Modified Newtonian Dynamics (MOND; \citealt{Milgrom1983mond}, see \citealt{Zhao2013mond, Banik2018mond, Bilek2018mond}).

In the context of satellite planes, galaxy mergers are generally raised as a potential explanation in terms of the tidal features they generate \citep[e.g.][]{Muller2018casp}. None the less, a more straightforward consequence may also contribute to the formation and enhancement of correlated planes of satellites -- a merger imprinting its angular momentum onto the participating satellite population, giving rise to a common rotational motion. 
\citet{Smith2016merger} (hereafter \citetalias{Smith2016merger}) set up major mergers in an $N$-body simulation between Milky Way or M31-mass haloes, with corresponding mass ratios of $1/2$ or higher. They reported that, given certain initial conditions were met, mergers could sling out satellites to large radii while their distribution along the axis (the \emph{interaction pole}) orthogonal to the plane in which the merger occurs (the \emph{interaction plane}) was maintained. This mechanism was capable of forming planar satellite distributions that were thin ($10-40\,\mathrm{kpc}$), extended ($150\,\mathrm{kpc}$), and rotationally stable over at least 6 Gyr.

\citetalias{Smith2016merger} reported that three initial conditions were especially conducive to correlated planes.
The satellites must inhabit a small range along the interaction pole -- as satellites are slung outwards along the interaction plane over the course of the merger, this initial orthogonal distribution defines the height of the resulting plane-of-satellites.
In order to maximise the angular momentum imparted onto their respective satellite populations, the two merging host galaxies must approach each other in near-circular trajectories.
Finally, the satellites' initial velocities must approximately align with the interaction plane, as any orthogonal component can weaken the longevity of the flattened satellite distribution. Dwarfs with prograde velocities with respect to the merger in-spiral are slung out to further distances, while retrograde dwarfs enter highly radial orbits and are tidally disrupted, resulting in a co-rotating system.

On the other hand, \citetalias{Smith2016merger}'s merger model is highly idealised. The more massive, \emph{primary} halo is assumed to be devoid of substructure, while the less massive, \emph{secondary} halo only contains 10 subhaloes -- the low sample size of which naturally results in an enhanced degree of flattening \citep[e.g.][]{Pawlowski2014pair}.
For merger mass ratios markedly below $1/1$, the primary halo should contain a larger satellite population that would be less-affected by angular momenta imparted by the secondary halo than vice-versa. Consequently, two populations are expected to co-exist in the merged system -- a thin plane consisting of less than half of the total satellite population, and a less-correlated distribution originating from the primary halo. This is at odds with the richly populated planes observed in the Local Volume.
Alternatively, a merger may occur between two host galaxies of similar mass, wherein satellite populations from both participating haloes should contribute equally in forming a correlated structure. This scenario would require both initial satellite distributions to be constrained along the interaction pole.
It is unclear -- given the stringent initial conditions required -- whether such a mechanism would be sufficiently commonplace in a full cosmological context.

And yet, the proposed histories of Local Volume hosts motivate the formation of satellite planes through major mergers as a interesting line of inquiry.
The Milky Way is thought to have followed a relatively quiescent evolutionary path after a $1/2.5$ merger with Gaia-Sausage-Enceladus (GSE; e.g. \citealt{Naidu2021gse}) $8-10$ Gyr ago. 
Conversely, M31 may have experienced a $1/3$ merger 5 Gyr ago \citep{Hammer2010tdg, Fouquet2012merger} and/or 2 Gyr ago with a M32 progenitor \citep{DSouza2018merger, Hammer2018merger}. The GPoA is thin (around $13\,\mathrm{kpc}$) and comprises 15 out of 27 satellites \citep{Ibata2013m31, Conn2013m31} -- the clear distinction between on-plane and off-plane satellites may arise from the comparatively low mass ratio of M31's last major merger. 
Centaurus A's disturbed morphology is argued to be a result of a $1/1.5$ merger 2 Gyr ago \citep{Wang2020merger} -- the CASP is thick (around $130\,\mathrm{kpc}$; \citealt{Muller2019casp}) and richly populated, which may be a consequence of a high-mass ratio merger event.
\citetalias{Smith2016merger}'s model may also account for the significant co-rotation observed in the VPOS \citep{Pawlowski2020dr2} and inferred from radial velocities from the GPoA \citep{Ibata2013m31} and CASP \citep{Muller2021casp}.

In this paper, we assess the imprint of recent major merger events between host galaxies on their present-time satellite distributions in high-resolution cosmological simulations.
We emphasize that while we frequently refer to \citetalias{Smith2016merger}'s model as a motivation for this work, we do not directly test the model itself and its corresponding conditions for satellite plane formation, and rather focus on the statistical impact of recent mergers on satellite planes.

This paper is structured as follows. In Section~\ref{sec:s2}, we sample simulated satellite systems from IllustrisTNG and define our metrics for quantifying any phase-space correlations present. In Section~\ref{sec:s3}, we search for the statistical imprint of recent merger events on the correlation of the most massive satellites at present time. We discuss the effect of post-merger satellite infall in Section~\ref{sec:s4}, and examine an alternate sample of satellites which was bound to their two merging host galaxies since the merger's beginning. We conclude in Section~\ref{sec:s5}.

\section{Systems at Present Time}
\label{sec:s2}

\subsection{Sampling Systems}
\label{sec:s2_sampling}

Systems of host galaxies and their satellite populations are drawn from the IllustrisTNG suite of cosmological simulations \citep[e.g.][]{tng100paper1, tng100paper2} -- specifically, the hydrodynamic TNG100-1 run. With a simulation box length of $L=110.7\,\mathrm{Mpc}$ at $z=0$ and a particle resolution of $M_{\mathrm{dm}}=7.5\times10^{6}M_{\odot}$, it provides a sufficiently populated statistical sample of hosts while resolving low-mass satellite subhaloes.

We sample haloes with virial masses within $M_{200}=[1,12]\times10^{12}M_{\odot}$. The lower mass bound is motivated by the halo masses of the Milky Way \citep{Posti2019mw} and M31 \citep{Chemin2009m31, Kafle2018m31}, while the upper bound is set by the commonly-adopted mass range for Centaurus A analogs \citep[e.g.][]{Muller2018casp, Muller2021casp, Pearson2022cena}.
In addition, we impose an isolation criterion by rejecting haloes with a companion with a mass $>0.25M_{200}$ within $3R_{200}$. This translates to around $800\,\mathrm{kpc}$ for Milky Way and M31-mass haloes, analogous to the distance between the Local Group hosts \citep{Li2021m31}, or around $1200\,\mathrm{kpc}$ for Centaurus A-mass haloes (which is comparable to the Centaurus A -- M83 separation; \citealt{Tully2015groups}).

For each halo and corresponding central galaxy, we proceed to identify a sample of 25 subhaloes as satellites. This population size roughly equates to the number of confirmed satellites with distances and radial velocities in M31 \citep{Conn2013m31} and Centaurus A \citep{Muller2021casp}, and is maintained for all halo masses to avoid systematic bias -- in general, a smaller satellite population with the same degree of phase-space correlation will appear to be more flattened, to a limit of $c/a=0$ at 3 satellites. \citep[see][]{Pawlowski2014pair}. Indeed, our choice of satellite count is ultimately arbitrary, only chosen to strike a balance between the number of sampled systems (which decreases with higher $N_{\mathrm{sat}}$) while retaining a sufficiently high number of satellites per system -- it is only crucial that $N_{\mathrm{sat}}$ be held constant. We also test a smaller $N_{\mathrm{sat}}=15$ in the \emph{low} sample (see Table~\ref{tab:s2_samplesize}) but find that apart from a systematically smaller $c/a$, all trends reported in the remainder of the paper for the \emph{full} sample also holds here.

Subhaloes recognized as gravitationally bound to the host halo in \texttt{Subfind} \citep{Springel2001subfind} are ranked by $M_{\mathrm{dm}}$ and a selection cut is made at 25 subhaloes.
Satellites are also required to each have a Main Progenitor Branch (MPB) merger tree constructed by \texttt{Sublink} \citep{Rodriguez-Gomez2015sublink}, corresponding to a history of more than one simulation snapshot. We henceforth refer to this initial sample as the \emph{full} satellite sample.

\begin{table*}
	\centering
	\caption{A summary of all key metrics defined and used in this study.}
	\vspace*{-1mm}
	\begin{tabular}{lllll}
		\hline
		Metric & Unit & Range & Description & Introduced In\\
		\hline
		$c/a$ & -- & $[0, 1]$ & Minor-to-major axis ratio (scale-free measure of plane flattening) & Section~\ref{sec:s2_psc} \\
		$\Delta_{\mathrm{rms}}$ & kpc & $[0, -]$ & Absolute root-mean-square plane height & Section~\ref{sec:s2_psc} \\
		$d_{\mathrm{rms}}/R_{200}$ & -- & $[0, -]$ & Satellite radial extent normalised to halo size & Section~\ref{sec:s2_psc}\\
		$\theta_{\mathrm{plane}}$ & degrees & $[0, 90]$ & Orbital pole concentration with respect to the best-fitting plane normal & Section~\ref{sec:s2_psc}\\
		$\theta_{\mathrm{orbit}}$ & degrees & $[0, 90]$ & Orbital pole concentration with respect to the best-fitting orbital pole & Section~\ref{sec:s2_psc}\\
		$N_{\mathrm{corr}}$ & -- (out of 25) & $[13, 25]$ & Number of co-rotating satellites with respect to the best-fitting plane normal & Section~\ref{sec:s2_psc} \\
		$t_{\mathrm{lookback}}$ & Gyr & $[0, 10]$ & Number of simulation snapshots from present time until end of last major merger & Section~\ref{sec:s3_identify} \\
		$\mu_{\mathrm{dm}}$ & -- & $(0, 1]$ & Mass ratio of merging central subhaloes in $M_{\mathrm{dm}}$ & Section~\ref{sec:s3_identify}\\
		$\Delta t_{\mathrm{merger}}$ & Gyr & $(0, 6]$ & Total duration of merger event & Section~\ref{sec:s3_identify}\\
		$h_{\mathrm{origin}}$ & $\mathrm{kpc}\,\mathrm{km}\,\mathrm{s}^{-1}$ & [0, --] & Total specific angular momentum of merging systems at said merger's beginning & Section~\ref{sec:s3_trajectory} \\
		$\Delta h$ & degrees & $[0, 45]$ & Root-mean-square spread of merger interaction poles (angular momenta vectors) & Section~\ref{sec:s3_trajectory}\\
		$\angle_{\mathrm{h, plane}}$ & degrees & $[0, 90]$ & Angle between mean interaction pole and best-fitting plane normal at $z=0$ & Section~\ref{sec:s3_trajectory}\\
		$\angle_{\mathrm{h, orbit}}$ & degrees & $[0, 90]$ & Angle between mean interaction pole and best-fitting satellite orbital pole at $z=0$ & Section~\ref{sec:s3_trajectory} \\
		$t_{\mathrm{infall}}$ & Gyr & $[0, 10]$ & Lookback time at which a satellite is first bound to its current host halo & Section~\ref{sec:s4_infall} \\
		$N_{\mathrm{part, end}}$ & satellites & $[0, 25]$ & Satellites bound to one of two merging haloes before the merger's end at $t_{\mathrm{lookback}}$ & Section~\ref{sec:s4_infall} \\
		$N_{\mathrm{part, origin}}$ & satellites & $[0, 25]$ & Satellites bound to one of two merging haloes before the merger's beginning & Section~\ref{sec:s4_infall} \\
        \hline
	\end{tabular}
	\label{fig:s2_definitions}
\end{table*}

Many of the obtained satellite populations are highly extended (central panel in Fig.~\ref{fig:s2_sample}), with a median distance beyond their hosts' virial radii at $1.03R_{200}$. There appears to be no correlation with system size.

We briefly compare the simulated satellite radial distribution with that observed for Local Group systems, adopting $R_{200,\mathrm{MW}}=200\,\mathrm{kpc}$ \citep{Wang2020mwmass} and $R_{200,\mathrm{M31}}=200\,\mathrm{kpc}$ \citep{Tamm2012m31}.
The simulated systems are somewhat consistent with the M31 satellites \citep{Savino2022m31}, whereas the Milky Way satellites \citep{McConnachie2012lg} are substantially more radially concentrated. We briefly check whether the latter is caused by a larger sample of fainter dwarfs (only observable due to their proximity) by limiting the Milky Way system to satellites with an absolute visual-band magnitude of $M_V \leq -6.4$, a threshold corresponding to the faintest satellite in the M31 sample. However, this magnitude-limited sample does not noticeably improve the Milky Way satellites' consistency with the simulated systems' radial distributions, and we obtain a similar result when only considering the 11 so-called classical satellites.

Extended satellite radial distributions have been noted to occur when only considering resolved subhaloes \citep{Riggs2022orphans, Sawala2023planes}, and a full treatment of artificially disrupted "orphan" satellites in a higher-resolution volume may be necessary to bring the simulated radial distribution in line with that observed for the Local Group satellite systems \citep[e.g.][]{Engler2021tng50}. However, we emphasise that this work focuses on the generalised effect of mergers on satellite distributions, and the above comparison with the Local Group hosts is made only for reference. Furthermore, the comparison is strongly dependent on the $R_{200}$ assumed for the Milky Way and M31, and a thorough consistency check is beyond the scope of this paper (and would be better performed by using absolute distances along with a better-constrained analog halo mass range). To compensate for the apparently radially extended nature of the simulated satellites, we additionally explore a separate sample of the 25 most massive satellites within the virial volume (the \emph{virial} sample) -- resulting in a truncated distribution as seen in the right-hand panel in Fig.~\ref{fig:s2_sample}. This alternate selection criterion is accompanied by a loss of around 50 per cent of Milky Way and M31-mass hosts ($M_{200}=[1,2]\times10^{12}\,M_{\odot}$) in the left-hand panel due to an insufficient number of viable satellites per halo, which results in a more uniform distribution of halo masses within our sampling range.

It has also been reported that the most luminous satellites do not necessarily trace the full underlying distribution of dark substructure, but of subhaloes with the most massive progenitors \citep{Libeskind2005cdm}. To account for any bias due to this discrepancy, we also trace the main progenitor branch of the 100 most massive satellites at present time and rank them by their maximum past dark matter mass (the \emph{progenitor} sample). Population sizes of the samples discussed above can be found in Table~\ref{tab:s2_samplesize}, along with their corresponding sampling criteria.

Finally, we take the positions and velocities of each \emph{full}-sample satellite relative to the central subhalo of the halo to which they are bound, and their magnitudes are then reassigned to unit vectors drawn randomly from an isotropic distribution. This isotropic sample serves as an null-hypothesis to compare simulated systems with while controlling for bias from radial distance profiles.

\begin{figure*}
	\includegraphics[width=0.311\textwidth]{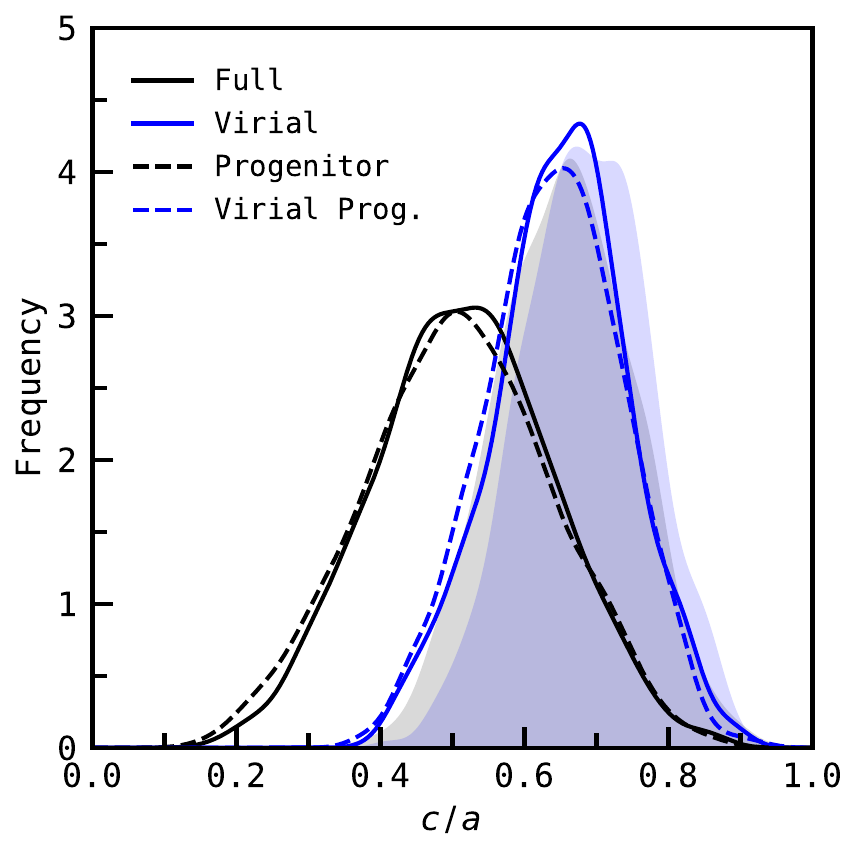}
	\includegraphics[width=0.329\textwidth]{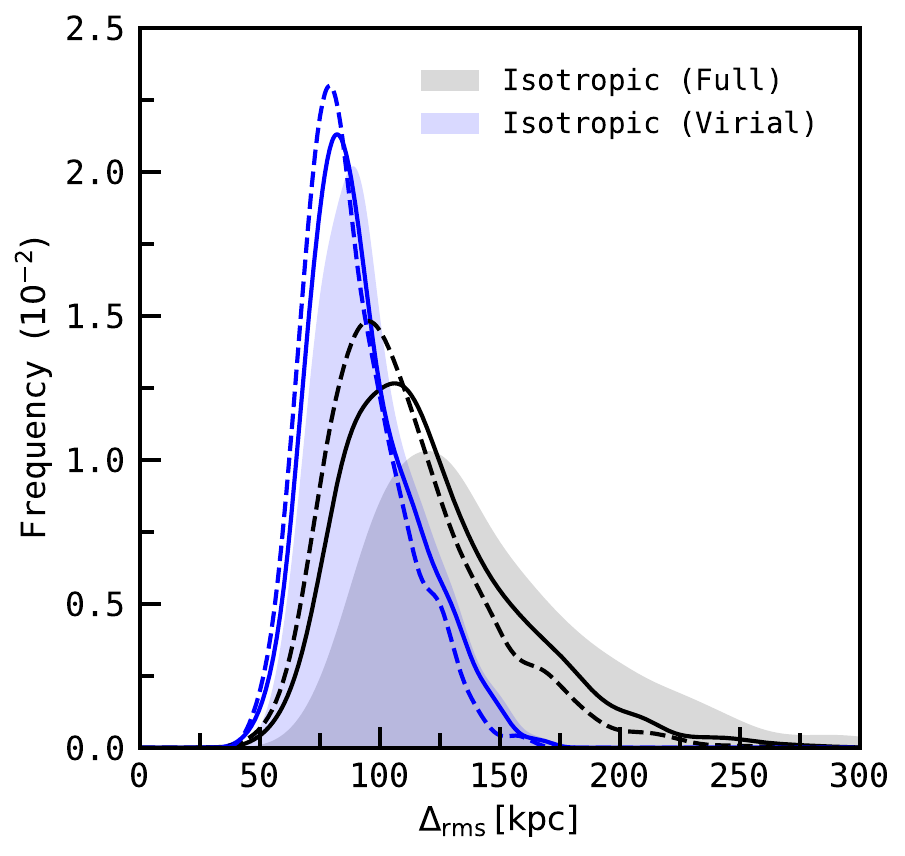}
	\includegraphics[width=0.31\textwidth]{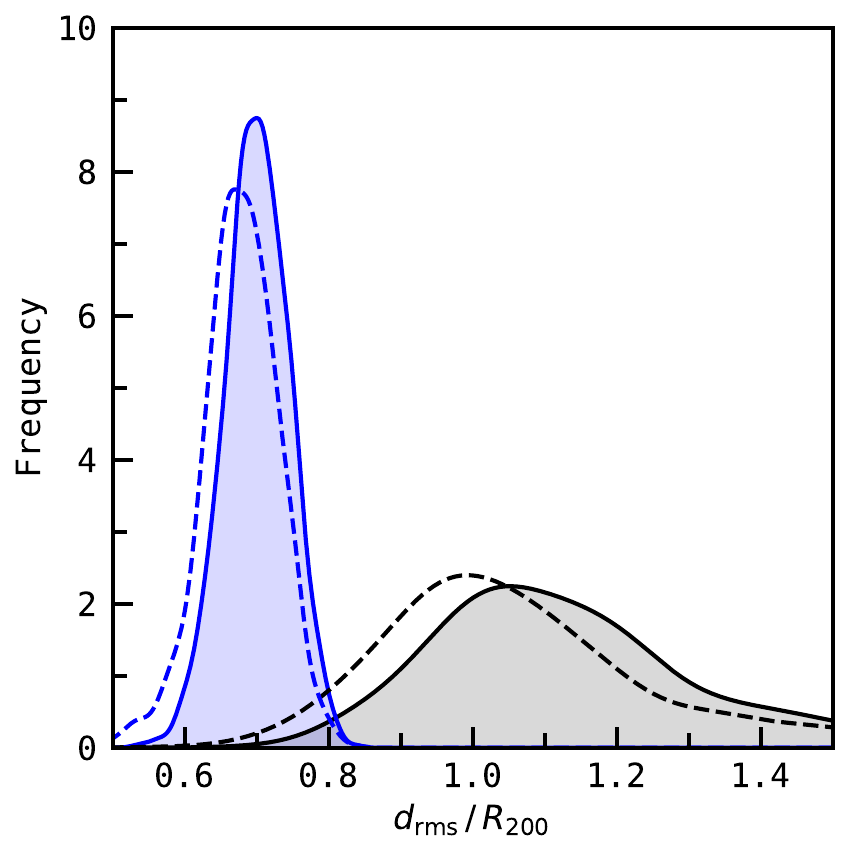}
	\includegraphics[width=0.32\textwidth]{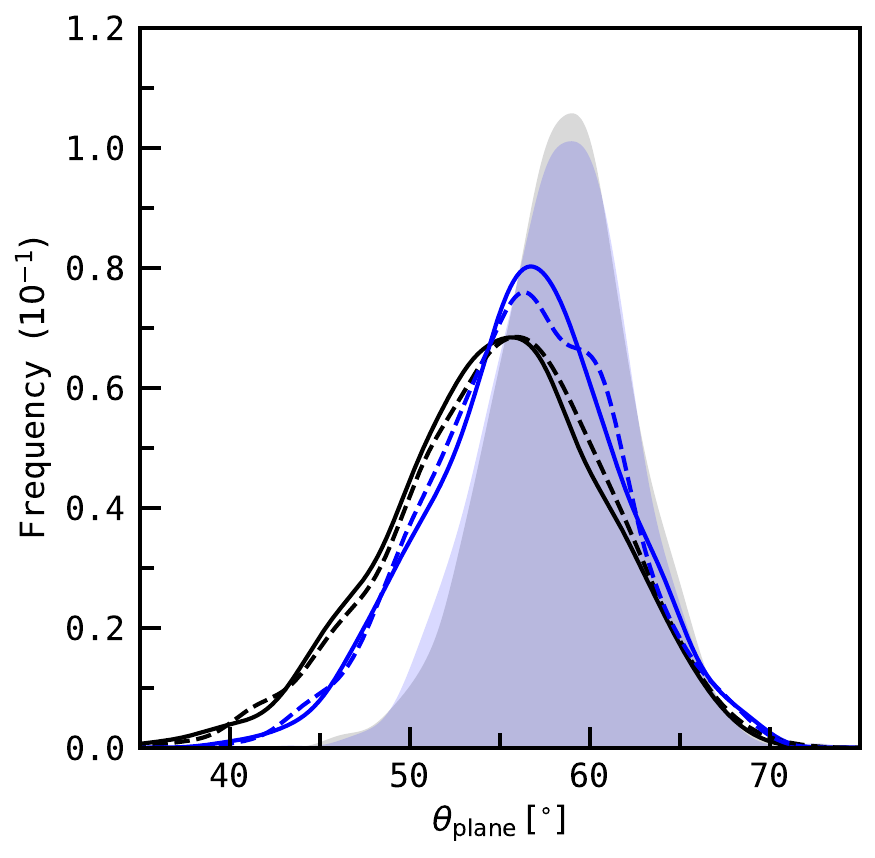}
	\includegraphics[width=0.32\textwidth]{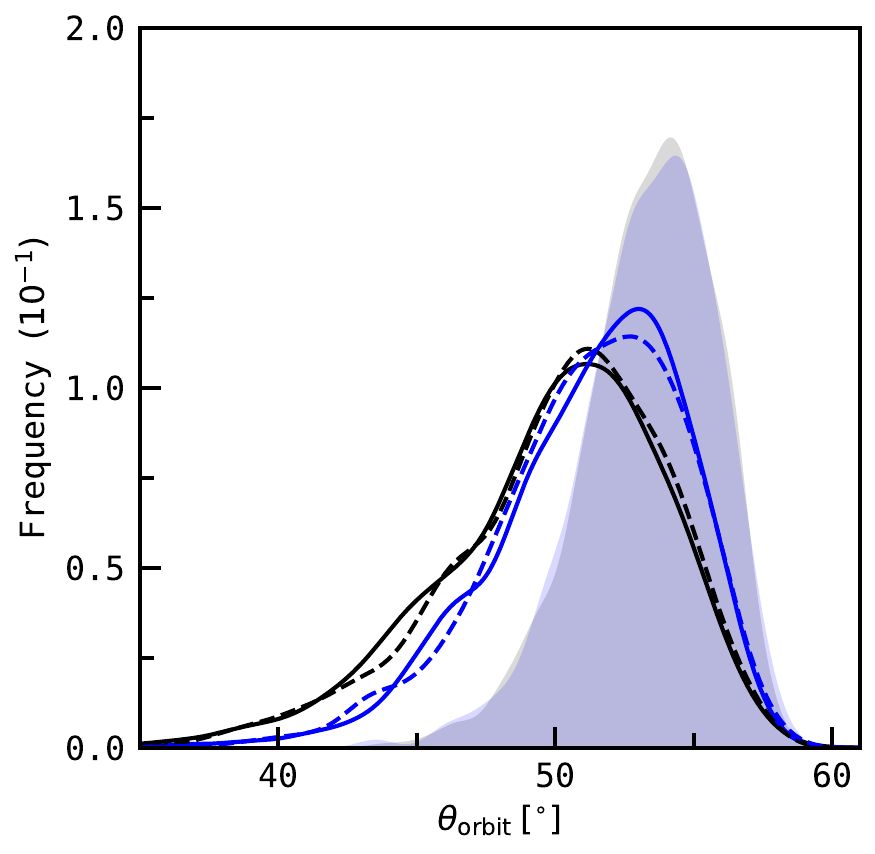}
    \includegraphics[width=0.32\textwidth]{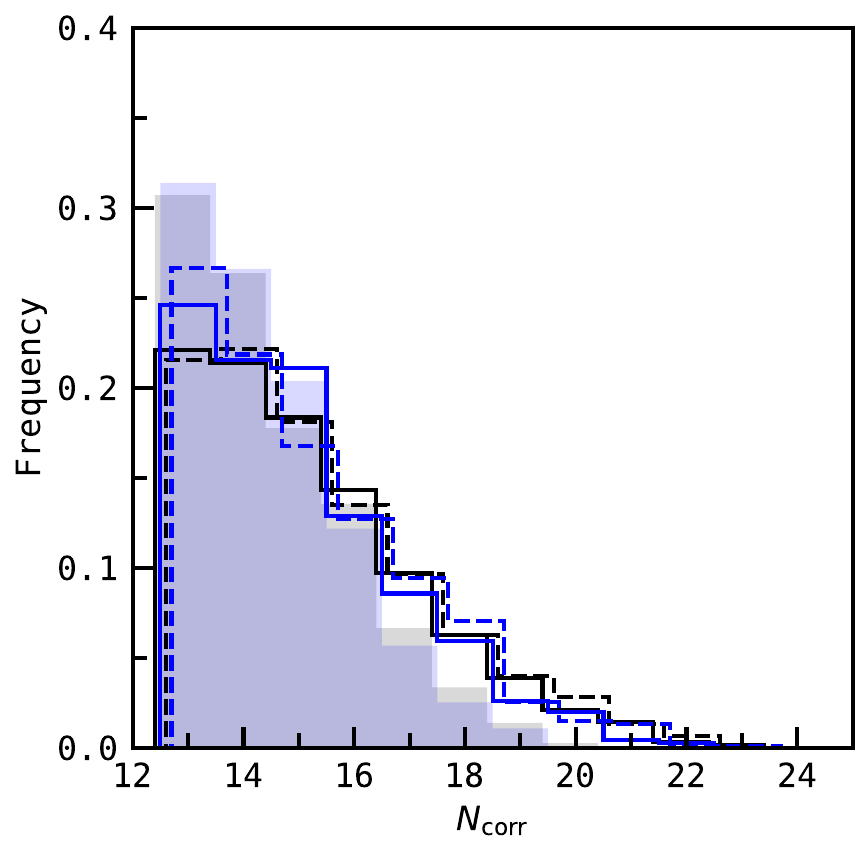}
    \caption{
    Present-time \emph{full}-sample distributions of spatial (upper row) and kinematic (lower row) phase-space correlation metrics, smoothed via Gaussian kernel density estimation. $N_{\mathrm{corr}}$, a discrete parameter, is shown as a histogram with shifted bins for visibility and is given out of a total of 25 satellites. Simulated systems are drawn in black or blue depending on whether their satellites were selected from the entire population of bound subhaloes (the \emph{full} sample) or restricted to within the virial volume (the \emph{virial} sample) respectively. The distribution of systems with satellites selected by their mass at present time are plotted by solid lines, whereas those chosen by their maximum mass along their main progenitor branch are drawn by dashed lines. Shaded regions represent isotropic systems, for which satellite positions and velocities are randomised while maintaining the radial distribution of their respective simulated systems. As the chosen method of ranking satellites is mostly irrelevant (with the possible exception of a slightly modified radial distribution from the sampling of different satellites), only isotropic counterparts of the \emph{full} and \emph{virial} systems are shown.
    }
    \label{fig:s2_present}
\end{figure*}

\subsection{Measuring phase-space correlations}
\label{sec:s2_psc}

We quantify the strength of spatial anisotropies in the sampled satellite systems using the following, commonly adopted parameters. Descriptions for all of these, as well as any new parameters introduced later in this work, can also be found in Table~\ref{fig:s2_definitions}.

The minor-to-major axis ratio, or $c/a$, is a scale-free measure of spatial correlation in satellite systems within the range $[0, 1]$. A lower value represents a strongly flattened, plane-like distribution, while higher values indicate near-isotropic systems. This is obtained via an unweighted tensor-of-inertia (ToI) fit \citep[e.g.][]{Metz2007distribution, Pawlowski2014pair}. For $N$ satellites with positions $\bm{r}_i$ where $i=[1,N]$ and a host galaxy located at $\bm{r}_0$, the system's unweighted moment of inertia tensor $\mathbfss{T}_0$ is
\begin{equation}
    \mathbfss{T}_0 = \sum_{i=1}^N \left[ (\bm{r}_i - \bm{r}_0)^2 \cdot \mathbfss{I} - (\bm{r}_i - \bm{r}_0) \cdot (\bm{r}_i - \bm{r}_0)^T \right],
    \label{eq:s2_toi}
\end{equation}
where $\mathbfss{I}$ is the identity matrix. The three eigenvectors of $\mathbfss{T}_0$, $\lambda_1 \leq \lambda_2 \leq \lambda_3$, correspond to the major, intermediate, and minor axes of the satellite distribution. $c/a$ is then obtained as the ratio between the root-mean-square (rms) magnitudes of the minor and major axes. Since the unweighted ToI is sensitive to outlying systems at larger radii \citep{Pawlowski2015problems}, we also parametrize the radial extent of a system by its satellites' rms radial distance $d_{\mathrm{rms}}$, normalised by their host halo's virial radius $R_{200}$. The absolute thickness of the distribution is expressed in terms of its plane height $\Delta_{\mathrm{rms}}$, which is equivalent to the rms spread of satellites along the minor axis.

With regards to kinematics, we examine the rms spherical distance of satellite orbital poles, or the direction of their orbital angular momenta, from a given reference vector. Prograde and retrograde poles are considered to be equivalent unless explicitly discussing co-rotation, in which case we require orbital poles to have a positive $z$-component -- otherwise, the pole's negative equivalent is taken instead.
We adopt two such metrics: $\theta_{\mathrm{plane}}$, the rms angle of satellite poles from the minor axis of their spatial distribution, and $\theta_{\mathrm{orbit}}$, which takes the best-fitting orbital pole as the reference vector instead. For a given system, we isotropically generate 32,400 grid vectors on a unit sphere, and identify the grid vector with the minimum associated satellite pole spread as the best-fitting pole.

We also identify the number of satellites co-orbiting with respect to their distribution's minor axis, $N_{\mathrm{corr}}$. For each system, its satellites' orbital poles are projected along this reference vector and are categorised by whether the resulting dot product is positive or negative -- $N_{\mathrm{corr}}$ is defined as the size of the more populated category. From our set sample size of $N_{\mathrm{sat}} = 25$ satellites per system, this approach results in a range of $N_{\mathrm{corr}}=[13, 25]$.

\subsection{Differences between sampling methods}
\label{sec:s2_volume}

On comparing samples of satellites ranked by present mass and their maximum progenitor mass, we find slightly lower mean values of $\Delta_{\mathrm{rms}}$ (a shift of 10 kpc from the \emph{full} sample) and $d_{\mathrm{rms}}/R_{200}$ (shift of 0.07) in the latter (see Fig.~\ref{fig:s2_present}). Subhaloes with the most massive progenitors are generally accreted earlier than the most massive subhaloes today (see Section~\ref{sec:s4}), with a correspondingly longer time frame to be slowed by dynamical friction and fall into the central regions of their host halo. Furthermore, halo masses are expected to be lower in earlier accretion events, while their satellites accordingly lie on lower-energy orbits.
On the other hand, the two approaches towards ranking satellites produce statistically indistinguishable distributions for $c/a$ and all three kinematic metrics, with Kolmogorov-Smirnov (KS) $p$-values consistently above $0.1$ for both \emph{full} and \emph{virial} sample types. Satellites located within subhaloes with the most massive progenitors inhabit marginally smaller regions in space but are not necessarily more correlated than those hosted by the most massive subhaloes at present time. Consequently, we disregard the progenitor-mass approach towards satellite selection and focus on ranking satellites by present mass for the remainder of this paper.

Distributions of phase-space correlation metrics for our present-time satellite systems are displayed in Fig.~\ref{fig:s2_present}. We employ a density estimation method with a Gaussian kernel to visually differentiate between similar distributions.

\begin{figure*}
	\includegraphics[width=0.317\textwidth]{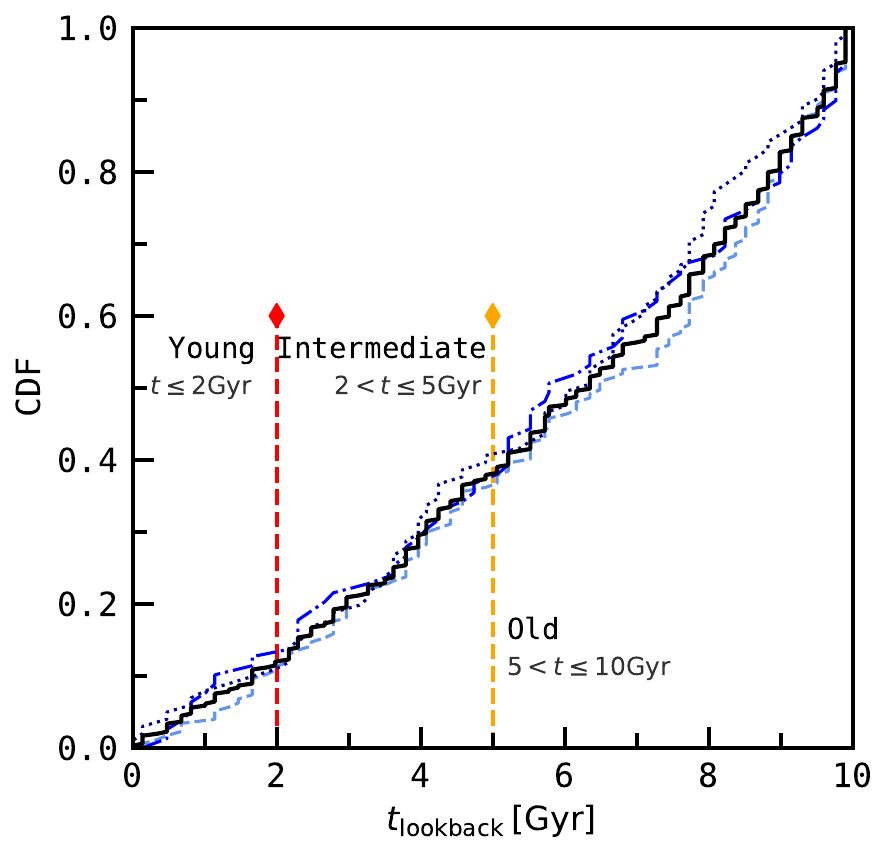}
	\includegraphics[width=0.32\textwidth]{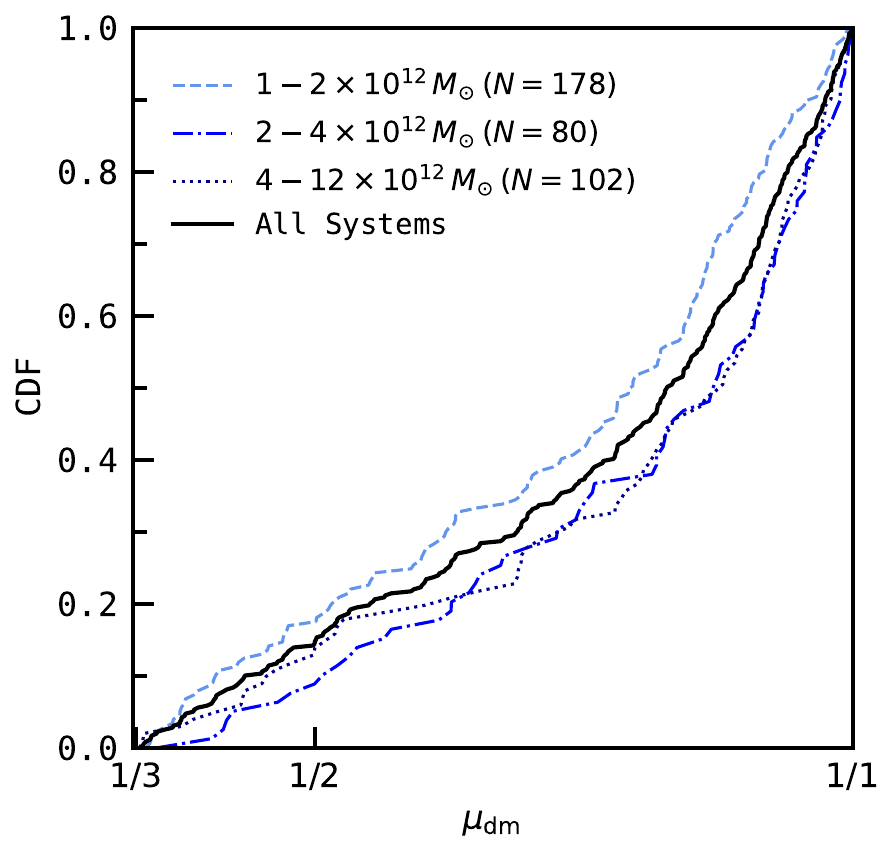}
	\includegraphics[width=0.315\textwidth]{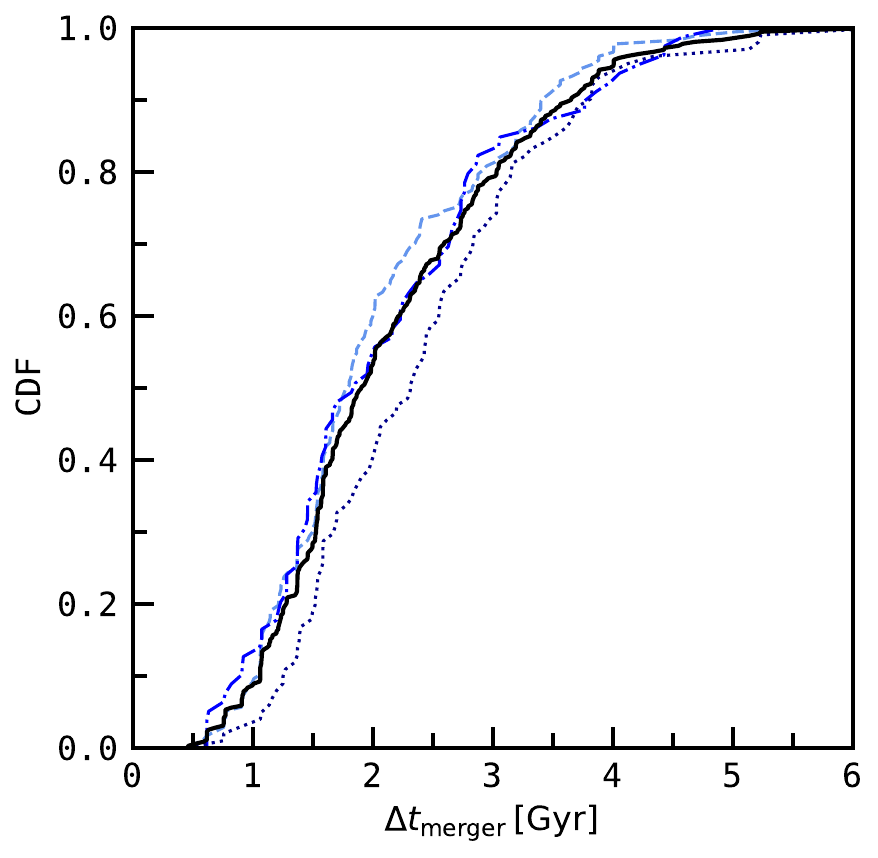}
    \caption{
    Properties of the last recorded major mergers experienced by simulated systems in the \emph{full} sample up to 10 Gyr ago. Systems are categorised by their halo mass (blue lines), whereas the black line indicates the net distribution of all \emph{full} systems.
    \textbf{Left Panel:} Lookback times to the mergers' completion, defined as when two subhaloes are recognized as one by \texttt{Sublink}. Time is discretized in snapshots in IllustrisTNG, but only physical time equivalents are shown here. Thresholds for \emph{young} systems ($0 \leq t_{\mathrm{lookback}} < 2\,\mathrm{Gyr}$) and \emph{intermediate} systems ($2 \leq t_{\mathrm{lookback}} < 5\,\mathrm{Gyr}$) are indicated with red and orange lines respectively.
    \textbf{Central Panel:} Merger mass ratios ($1/3 \leq \mu_{\mathrm{dm}} \leq 1$), determined by the maximum progenitor masses of the two merging subhaloes.
    \textbf{Right Panel:} Merger durations, defined as the time between when the two merging subhaloes are separated by a distance equivalent to the sum of their host haloes' virial radii and the merger's completion.
    }
    \label{fig:s3_mergerproperties}
\end{figure*}

\begin{table}
	\centering
	\caption{Criteria used when categorising systems by the recency of their last major merger event. Corresponding sample sizes of \emph{full} and \emph{virial} systems are shown under $N_{\mathrm{full}}$ and $N_{\mathrm{virial}}$ respectively.}
	\vspace*{-1mm}
	\begin{tabular}{lllll}
		\hline
		Type & Category & Last Merger (Gyr) & $N_{\mathrm{full}}$ & $N_{\mathrm{virial}}$ \\
		\hline
		Quiescent & -- & $t_{\mathrm{lookback}} > 10$ & 1131 & 813 \\
		Merger & Old & $5 < t_{\mathrm{lookback}} \leq 10$ & 220 & 170 \\
		" & Intermediate & $2 < t_{\mathrm{lookback}} \leq 5$ & 101 & 70 \\
		" & Young & $t_{\mathrm{lookback}} \leq 2$ & 41 & 35 \\
		\hline
	\end{tabular}
	\label{tab:s3_categories}
\end{table}

The degree of flattening inherent in our \emph{full} sample is distinctly stronger than in the \emph{virial} sample, with a mean $c/a$ difference of $0.13$. The \emph{virial} sample's $c/a$ distribution for simulated systems is very similar to that of its isotropic counterpart.
The isotropic \emph{full} and \emph{virial} systems each display a similar $c/a$ distribution, despite their dramatically different radial distributions (as is apparent in the upper-right panel of Fig.~\ref{fig:s2_present}) -- larger radial extents do not necessarily impact the flattening of isotropic distributions in a systematic manner.
These trends appear to suggest that simulated satellites are distributed in a near-isotropic manner within the virial volume, whereas -- as seen from the mean $\Delta_{c/a} = 0.15$ shift between \emph{full} simulated and isotropic systems -- a greater degree of anisotropy can be found at greater radii.
We also report a significant negative correlation between $d_{\mathrm{rms}}/R_{200}$ and $c/a$ in the \emph{full} sample with a corresponding Kendall rank correlation coefficient of $\tau = -0.34$, but a similar correlation is not present in the \emph{virial} sample ($\tau = 0.01$, $p = 0.56$, see Fig.~\ref{fig:aa_extent_ca}). 
The prominent flattening of systems in the \emph{full} sample appears to be primarily driven by this anisotropy in distant satellites, for which the presence of local filaments -- and triaxial halo morphologies -- is likely responsible.

\section{Merger Histories}
\label{sec:s3}

\subsection{Identifying major mergers}
\label{sec:s3_identify}

We now look to ascertain whether phase-space correlations in simulated satellite systems are enhanced if they experienced a major merger event with a host galaxy of comparable mass. Galaxy mergers in IllustrisTNG are identified by the \texttt{Sublink} algorithm \citep{Rodriguez-Gomez2015sublink} when the corresponding subhalo has more than one direct progenitor. Cases where three or more progenitors exist are treated as separate, binary mergers between the primary progenitor and each secondary subhalo.

We elect to define major mergers as mergers between two central subhaloes with a dark mass ratio of $\mu_{\mathrm{dm}} \geq 1/3$. This threshold, while arbitrary, nevertheless encompasses estimated mass ratios of the last notable mergers proposed for M31 \citep{DSouza2018merger, Hammer2018merger} and Centaurus A \citep{Wang2020merger}, as well as the Milky Way-GSE merger \citep{Naidu2021gse}, and is roughly consistent with definitions from \citet{Rodriguez-Gomez2015sublink} and \citet{Hopkins2010merger}.
As discussed in Section~\ref{sec:s1}, only the less massive, secondary halo should receive a strong angular momentum imprint in a low-mass ratio merger (\citetalias{Smith2016merger}) -- its correspondingly low satellite galaxy population would prevent the potential formation of a richly populated plane-of-satellites. In addition, lower-mass ratio mergers can be considered more akin to group infall, which is outside the scope of this work. As a result, we disregard mergers below this $1/3$ threshold for the remainder of this paper.

Before a merger event, mass is generally accreted from the secondary subhalo by the primary subhalo, resulting in the former retaining only a minimal amount of dark mass before the merger's completion. Mass ratios are consequently calculated from the maximum progenitor masses of both participating central subhaloes. For near-$1/1$ mergers, this occasionally causes a discrepancy between the direction of mass accretion and the maximum progenitor masses of the two subhaloes -- to avoid cases where $\mu_{\mathrm{dm}} > 1$, we flip the primary and secondary designations in such rare instances.

\begin{figure*}
	\includegraphics[width=0.304\textwidth]{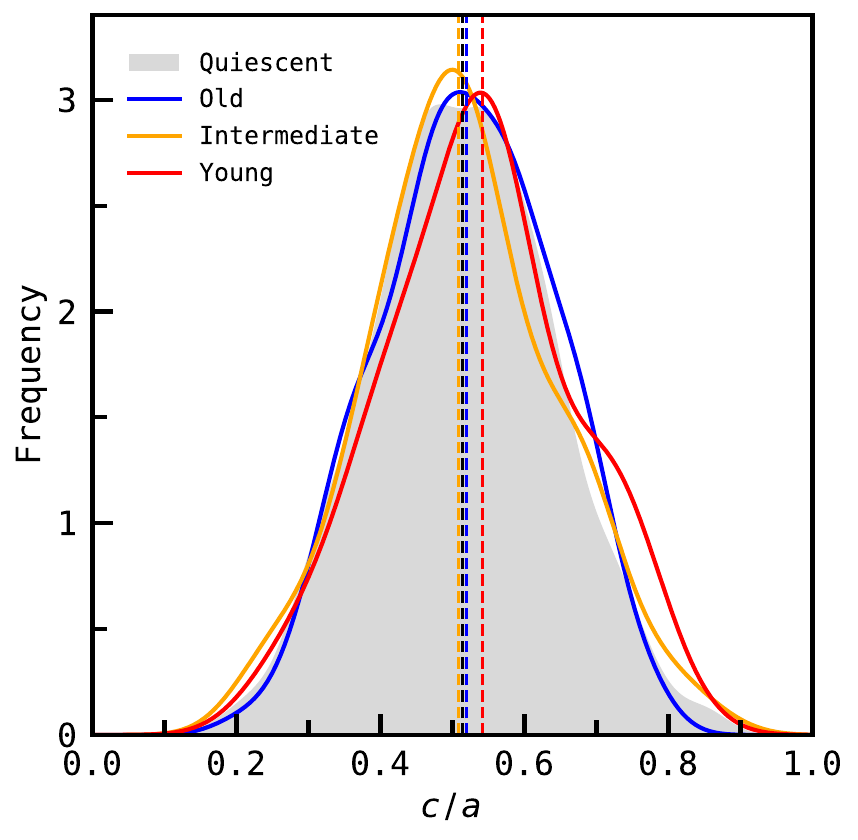}
	\includegraphics[width=0.321\textwidth]{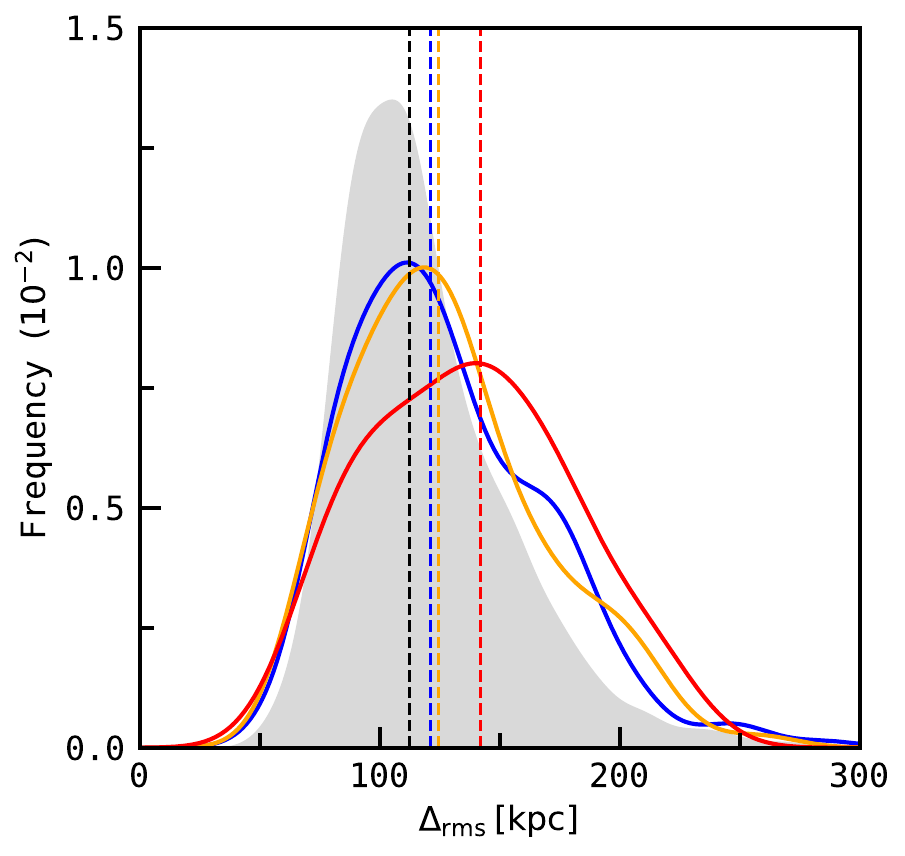}
	\includegraphics[width=0.311\textwidth]{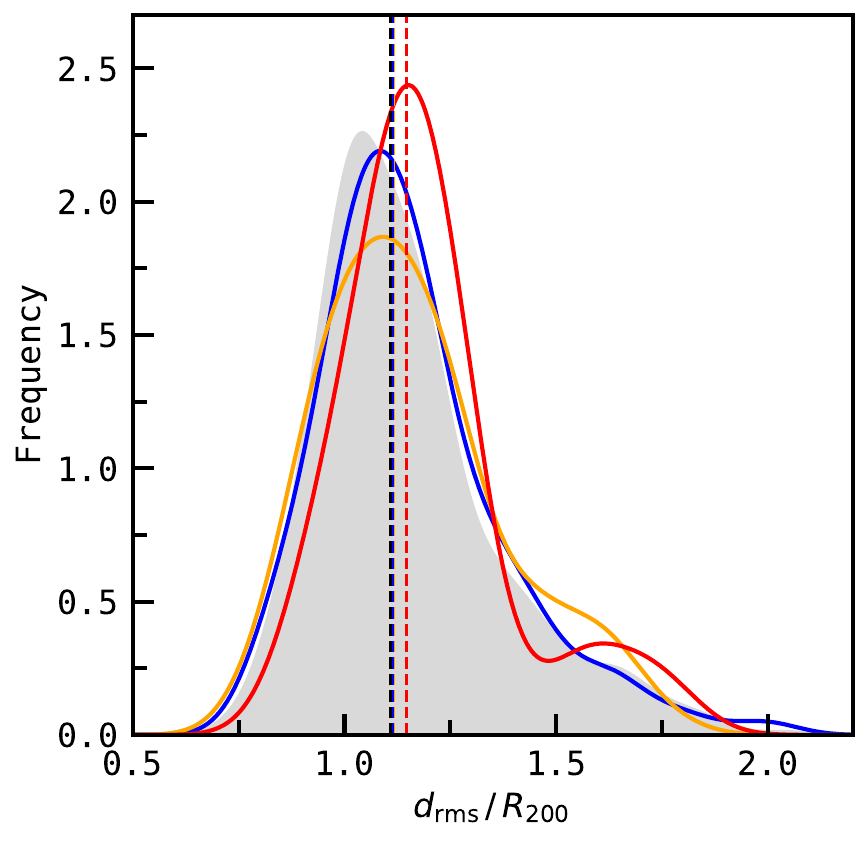}
	\includegraphics[width=0.304\textwidth]{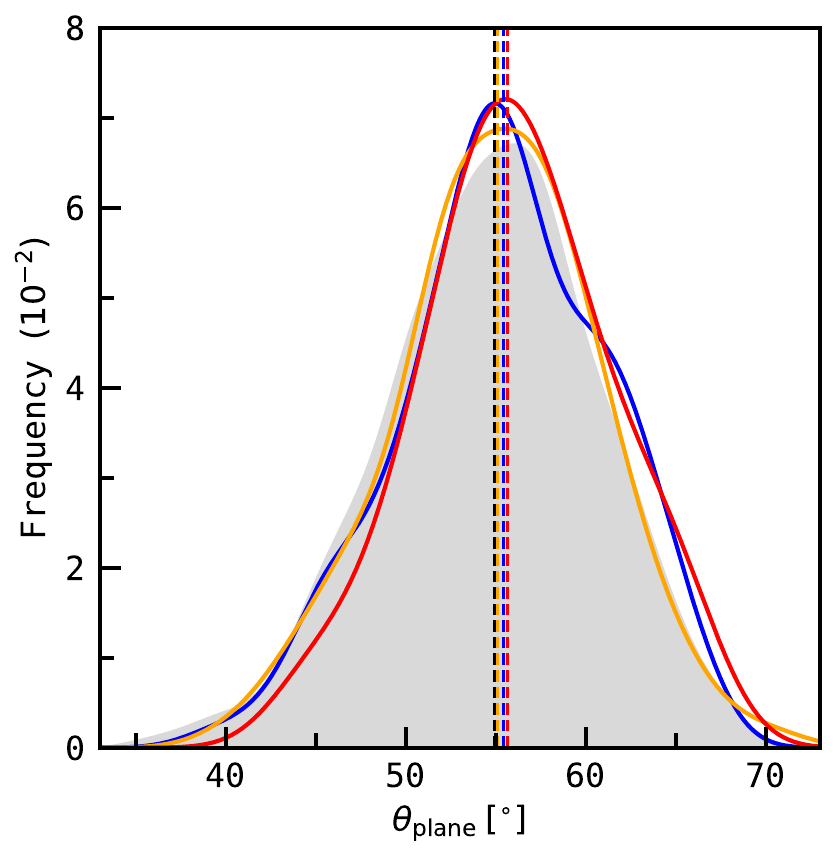}
	\includegraphics[width=0.321\textwidth]{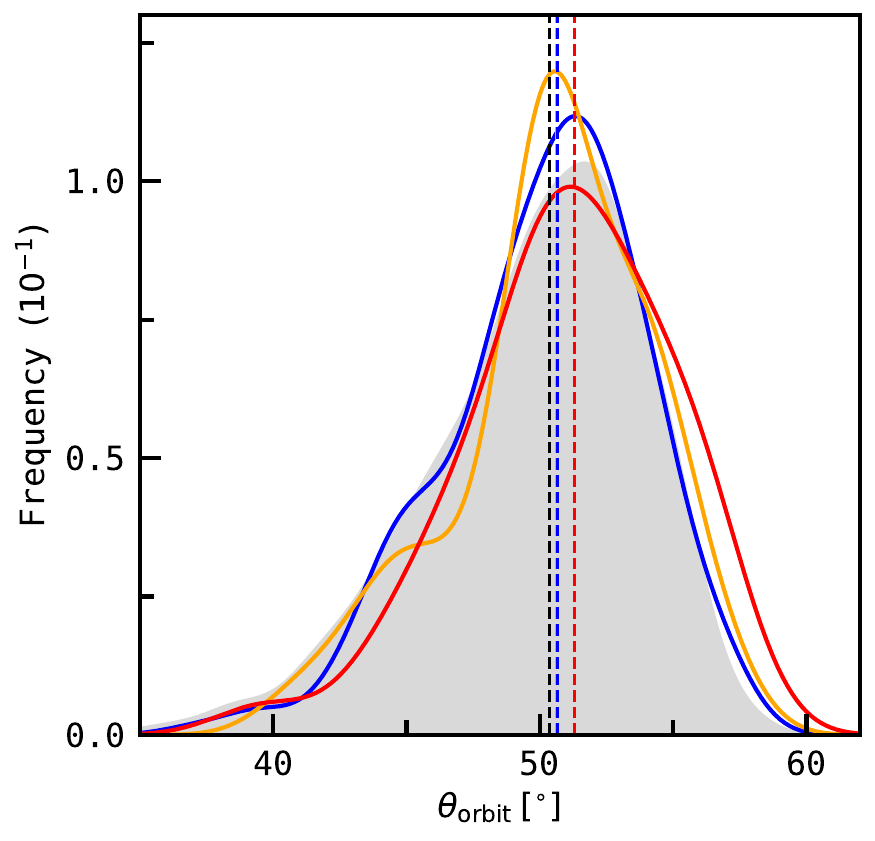}
	\includegraphics[width=0.320\textwidth]{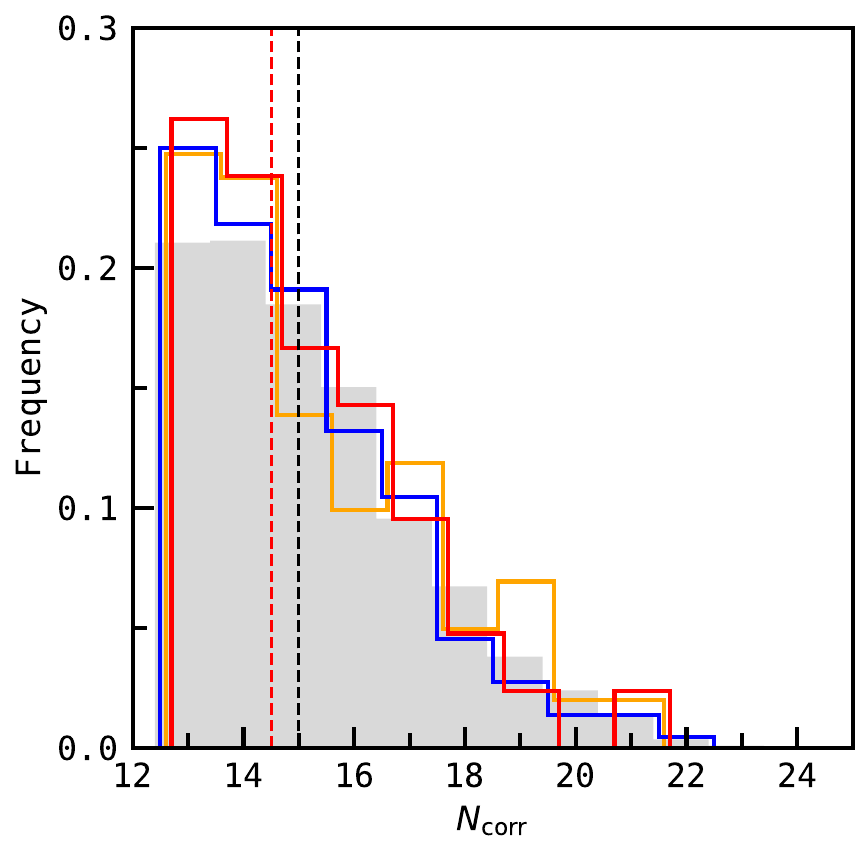}
    \caption{
    Present-time distributions of spatial (upper row) and kinematic (lower row) phase-space correlation metrics in the \emph{full} sample, categorised by the recency of each system's most recent major merger (see Table~\ref{tab:s3_categories} for definitions). Distributions are smoothed via Gaussian kernel density estimation -- $N_{\mathrm{corr}}$, a discrete parameter, is instead shown as a histogram with shifted bins for visibility. \emph{Quiescent-type} systems are shaded grey, whereas systems in the \emph{old}, \emph{intermediate}, and \emph{young} categories are drawn in blue, orange, and red respectively. Median values are indicated by dashed lines of the corresponding colour.
    }
    \label{fig:s3_byhistory}
\end{figure*}

We only consider the last major merger that systems have experienced, as each new event should imprint differently oriented angular momenta onto their satellites. We additionally reject systems with a second major merger within a lookback time of $0.5\,\mathrm{Gyr}$ of its last major merger. \citetalias{Smith2016merger} reported that stable planes require a similar time to form after the merger's completion -- systems that experience consecutive mergers may be too chaotic to recover the imprint of a individual event.

We categorise systems by the recency of their last major merger, as summarised in Table~\ref{tab:s3_categories}. We first divide the \emph{full} and \emph{virial} samples into \emph{merger-type} and \emph{quiescent-type} systems based on whether they have experienced a major merger within the last $10\,\mathrm{Gyr}$. Since we only trace subhalo merger trees for 10 Gyr, \emph{quiescent-type} systems lacking a major merger in their history cannot be used to compare the said merger's properties with present-time satellite distributions. Additionally, we subdivide \emph{merger-type} systems into \emph{old}, \emph{intermediate}, and \emph{young} systems based on $t_{\mathrm{lookback}}$ to their last merger's completion. Their maximum lookback time thresholds of $10\,\mathrm{Gyr}$, $5\,\mathrm{Gyr}$, and $2\,\mathrm{Gyr}$ are motivated by the estimated ages of the Milky Way-GSE, M31, and Centaurus A mergers respectively.

Fig.~\ref{fig:s3_mergerproperties} plots cumulative distributions of merger completion time $t_{\mathrm{lookback}}$, mass ratio $\mu_{\mathrm{dm}}$, and duration $\Delta t_{\mathrm{merger}}$. We also check for mass dependence by placing systems into three bins -- one corresponding to Milky Way and M31-mass systems, one intermediate bin, and one corresponding to Centaurus A-mass systems -- and plotting them separately. $t_{\mathrm{lookback}}$ does not have a statistically significant mass dependence (KS: $p > 0.25$) whereas less massive systems appear to tend towards mergers with lower mass ratios. Unsurprisingly, Centaurus A-mass systems demonstrate marginally longer merger durations as a natural consequence of their larger characteristic distances and time-scales.

\subsection{Comparison with merger history}
\label{sec:s3_imprint}

Fig.~\ref{fig:s3_byhistory} plots the distribution of our satellite phase-space correlation metrics when systems are categorised by their hosts' merger histories. We primarily consider \emph{full} systems here -- \emph{virial} systems risk disregarding satellites that happen to be slung outwards (as proposed in \citetalias{Smith2016merger}'s merger model) to beyond the virial volume.

The four distributions for each correlation metric are statistically indistinguishable except for $\Delta_{\mathrm{rms}}$, wherein systems that have experienced a merger in the last 2 Gyr host thicker satellite distributions.
While larger values of $c/a$ and $d_{\mathrm{rms}}/R_{200}$ can both contribute to such an increase in plane height, only radial extent appears to differ depending on merger history to a statistically significant degree (KS: $p=0.02$ between \emph{quiescent-type} and \emph{young} systems). We also recover a comparable shift in $\Delta_{\mathrm{rms}}$ and $d_{\mathrm{rms}}/R_{200}$ from the \emph{virial} sample, while the $c/a$ distribution holds constant with merger lookback time.

\begin{figure*}
	\includegraphics[width=0.325\textwidth]{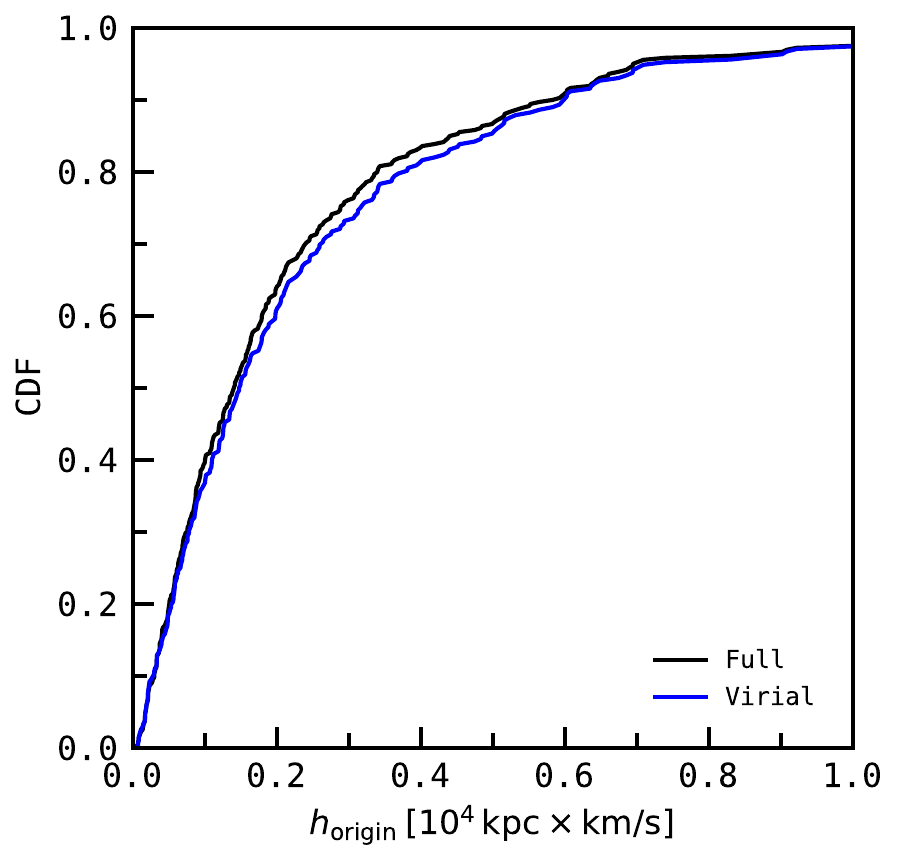}
	\includegraphics[width=0.319\textwidth]{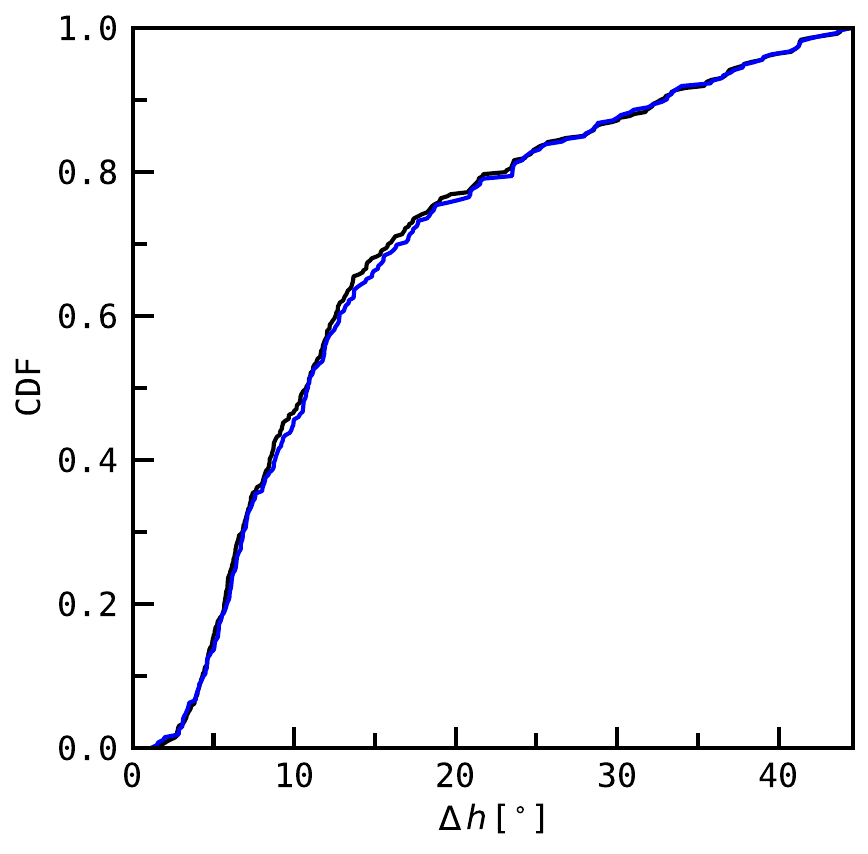}
	\includegraphics[width=0.32\textwidth]{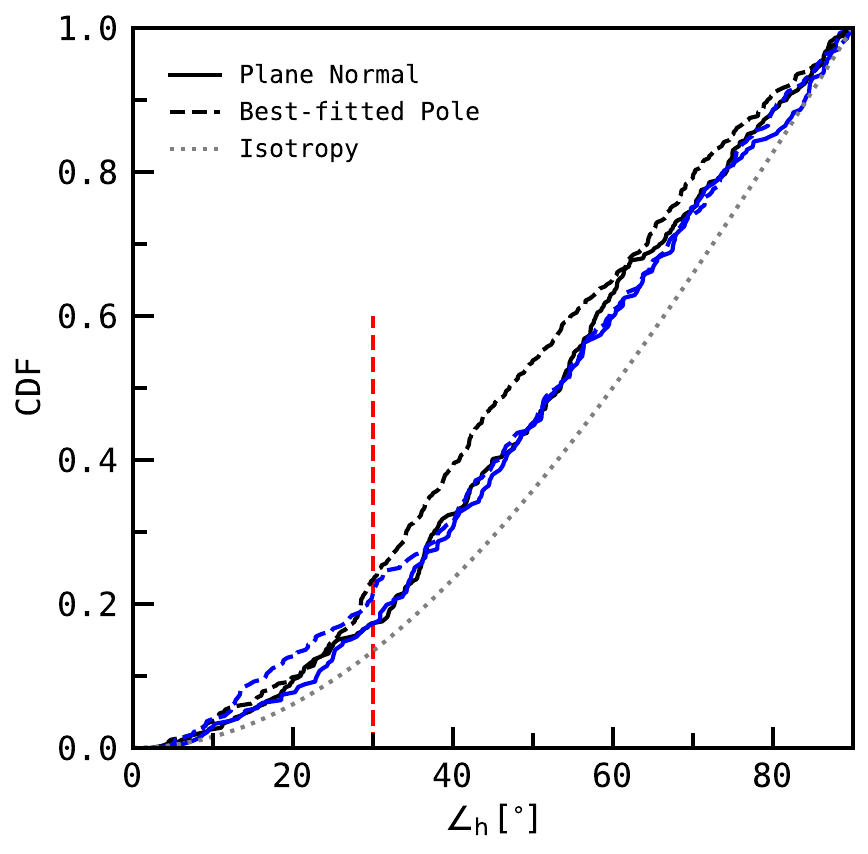}
    \caption{Cumulative distributions of interaction trajectory parameters. \textbf{Left Panel:} Summed specific angular momentum of each merging host galaxy pair. Results for \emph{full} and \emph{virial} samples are drawn in black and blue respectively. Higher values correspond to more circular interaction trajectories. \textbf{Central Panel:} The rms spread of interaction poles, or the direction of angular momentum of the merger in-spiral, averaged over the first 50 per cent of the merger's duration. Higher values imply a greater degree of instability. \textbf{Right Panel:} Angle between the mean interaction pole and the best-fitting plane normal or best-fitting satellite orbital pole at present time, indicated with solid and dashed lines respectively. The expected distribution of the alignment between two isotropically drawn vectors is shown as a dotted grey line.}
    \label{fig:s3_traj_dists}
\end{figure*}

However, the main driver of $\Delta_{\mathrm{rms}}$'s correlation with $t_{\mathrm{lookback}}$ appears to be a trend wherein more massive systems tend to have experienced a major merger more recently -- the median halo mass of \emph{young} systems is around 1.3 times larger than that of \emph{quiescent-type} systems (see Fig.~\ref{fig:aa_scale_byhistory}), with a corresponding difference in median $R_{200}$ of $23\,\mathrm{kpc}$. Since $d_{\mathrm{rms}}/R_{200}$ only reflects the radial extent of satellites with respect to their host halo, the marked increase in absolute plane heights with merger recency is accompanied only by lesser median shifts in $c/a$ and $d_{\mathrm{rms}}/R_{200}$ in Fig.~\ref{fig:s3_byhistory}.

$\Delta_{\mathrm{rms}}$ also appears to have a weak positive correlation with merger mass ratio in both samples (Kendall: $p=5.1\times10^{-3}$ \emph{full}, $p=4.4\times10^{-4}$ \emph{virial}), wherein systems with near-$1/1$ mergers tend to have thicker planes. A high mass ratio is also weakly correlated with a less-flattened satellite distribution in the \emph{virial} sample ($p=9.4\times10^{-3}$), but no significant correlation was found in their \emph{full}-sample counterparts.
Conversely, distributions of the three kinematic metrics are indistinguishable within the scope of statistical fluctuations. It is still interesting to note that \emph{recent} and \emph{active} systems demonstrate marginally less correlated satellite distributions than their more quiescent counterparts according to all phase-space correlation metrics -- as measured by their median values -- in both \emph{full} and \emph{virial} samples.
Regardless, it is evident that the presence of a recent major merger by itself does not generally strengthen the phase-space correlation of present-time systems. If anything, it appears to weaken the degree of correlation.

\subsection{Infall trajectories}
\label{sec:s3_trajectory}

Until now, we have focused only on the existence (or lack thereof) of major host mergers in the history of satellite galaxy systems, finding no significant enhancement in the phase-space correlation of those that have experienced recent merger events. From the apparent lack of highly flattened and kinematically correlated systems at present time -- especially so when restricting satellites to the virial volume -- it seems unlikely that coherent satellite distributions as observed in the Local Volume are reliably formed by any mechanism self-consistently included in IllustrisTNG.

Nevertheless, it is conceivable that mergers with specific initial conditions may improve satellite correlation, even if not necessarily to the observed extent. \citetalias{Smith2016merger} reported that mergers between host galaxies that follow circular infall trajectories were conducive to efficient angular momentum injection, slinging out satellites to greater distances and forming flattened distributions. In their N-body simulations, \citetalias{Smith2016merger} specify the nature of the two central subhaloes' infall trajectory by setting the initial tangential velocity as a fraction $f_{\mathrm{circ}}$ of the system's two-body orbital velocity. $f_{\mathrm{circ}}=1$ would imply a fully circular merger, whereas  $f_{\mathrm{circ}}=0$ represents a radial, head-on collision. However, their model lacks an initial radial velocity component, which would be expected in a cosmological context. 

We instead elect to describe merger trajectories by the summed magnitudes $h$ of the central subhaloes' specific angular momenta. Unlike $f_{\mathrm{circ}}$, our metric is scale-dependent and has no defined upper limit, but nevertheless serves as a rough tracer of a merger's infall trajectory -- large values of $h$ correspond to a more circular merger, whereas small values suggest a radial infall. Naturally, the angular momentum and corresponding tangential velocity of the central subhaloes are expected to decay over the course of a merger due to the effects of dynamical friction. We compensate for this by calculating $h$ in the early stages of the merger, which should be more representative of its initial interaction trajectory.

\citetalias{Smith2016merger} defined a merger's beginning as when the secondary, less massive central subhalo is located at the primary halo's virial radius. This choice is in accordance with their assumption that the primary halo lacks substructure -- we argue a merger instead between two richly populated haloes would have already progressed significantly at this stage due to their satellite populations overlapping. We instead define the merger \emph{origin} as the last snapshot before the distance between the merging central subhaloes is equivalent to the sum of their respective haloes' virial radii. If both are already bound to a single halo, we take double its virial radius as the threshold distance. Using this definition, we parametrize a merger's infall trajectory as specific angular momentum $h$ measured at its \emph{origin}, $h_{\mathrm{origin}}$.

Mergers in a cosmological context do not occur along a static interaction plane. External forces and torques will shift the interaction pole $\bm{h}$ over time, especially in an encounter's final stages. To quantify this instability, we measure the rms spread $\Delta h$ in $\bm{h}$ over the first half of a merger's duration. Note that the orientation of the interaction pole may flip as the distance between the system's barycentre and the two host galaxies shrinks in the merger's later stages, briefly reversing the rotational sense of the hosts with respect to their centre-of-mass. As such flips do not necessarily imply an unstable merger, reversed $\bm{h}$-vectors are considered identical to their original counterparts.

\begin{figure*}
	\includegraphics[width=0.31\textwidth]{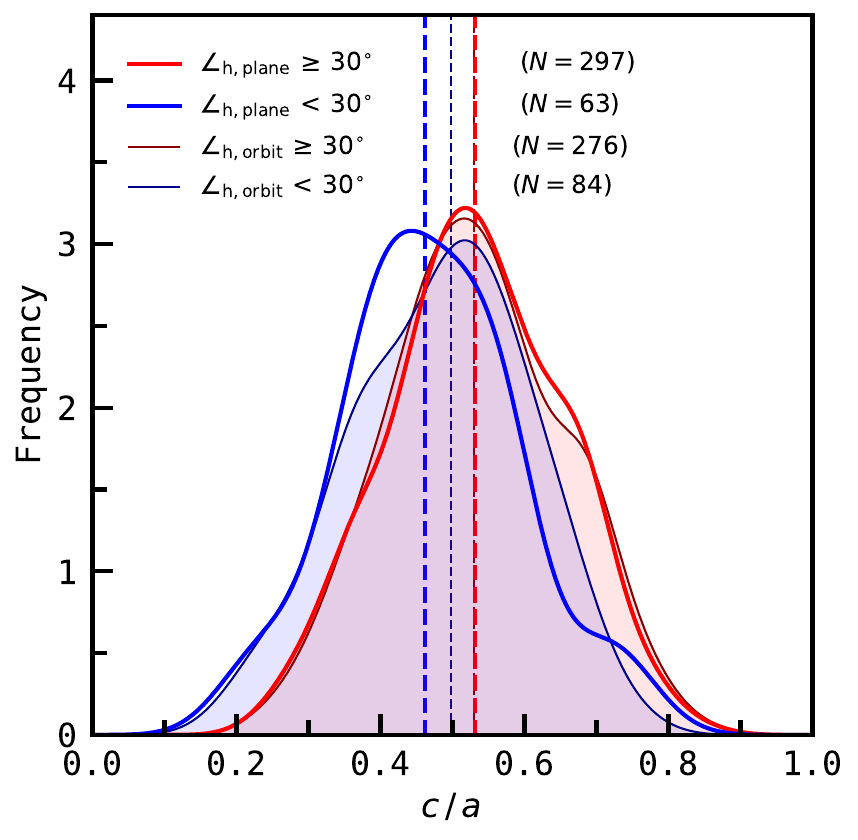}
	\includegraphics[width=0.325\textwidth]{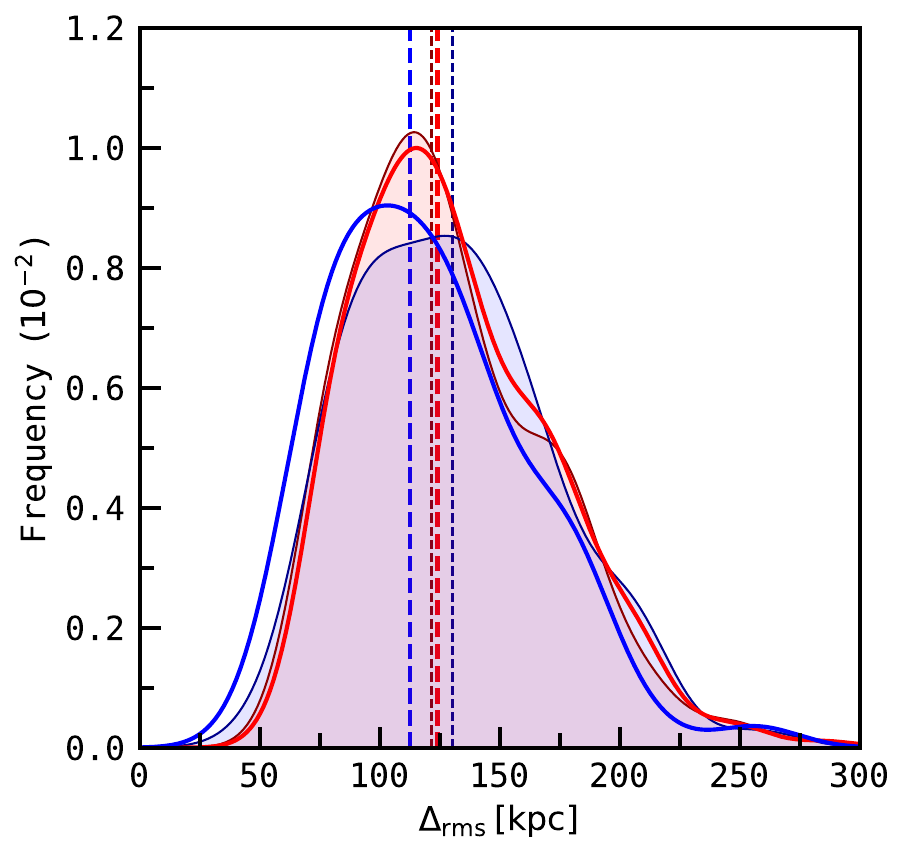}
	\includegraphics[width=0.322\textwidth]{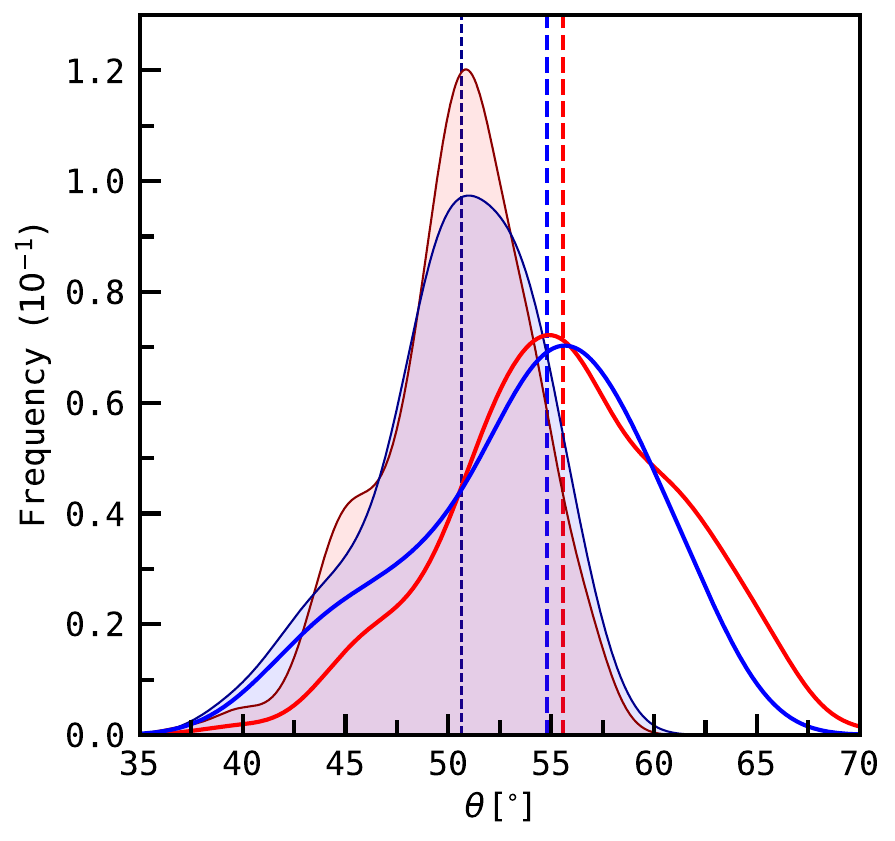}
    \caption{Minor-to-major axis ratio $c/a$, plane height $\Delta_{\mathrm{rms}}$, and orbital pole spread $\theta$ for present-time satellite systems in the \emph{full} sample, categorised by their best-fitting plane normal or best-fitting orbital pole's alignment with their last major merger's interaction pole (unshaded and shaded regions respectively). Well-aligned systems are drawn in blue, whereas less-aligned systems are drawn in red. Median values are indicated by dashed lines of the corresponding colour. Categorised distributions for other phase-space correlation metrics are statistically indistinguishable and are not displayed here for brevity.}
    \label{fig:s3_traj_kde}
\end{figure*}

\begin{figure*}
	\includegraphics[width=0.313\textwidth]{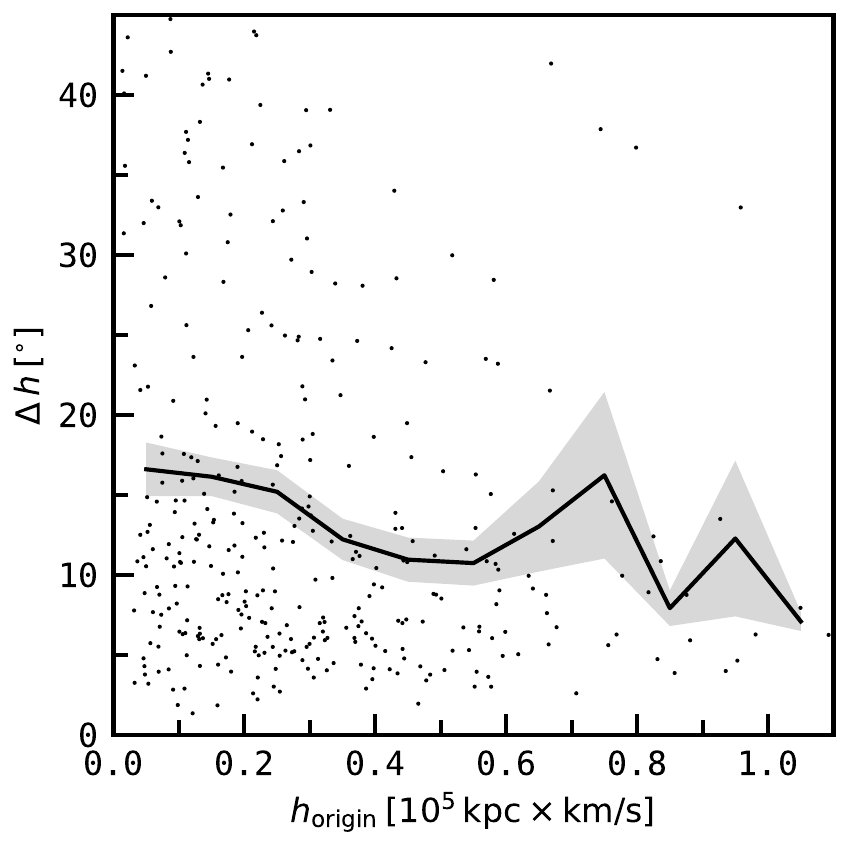}
	\includegraphics[width=0.315\textwidth]{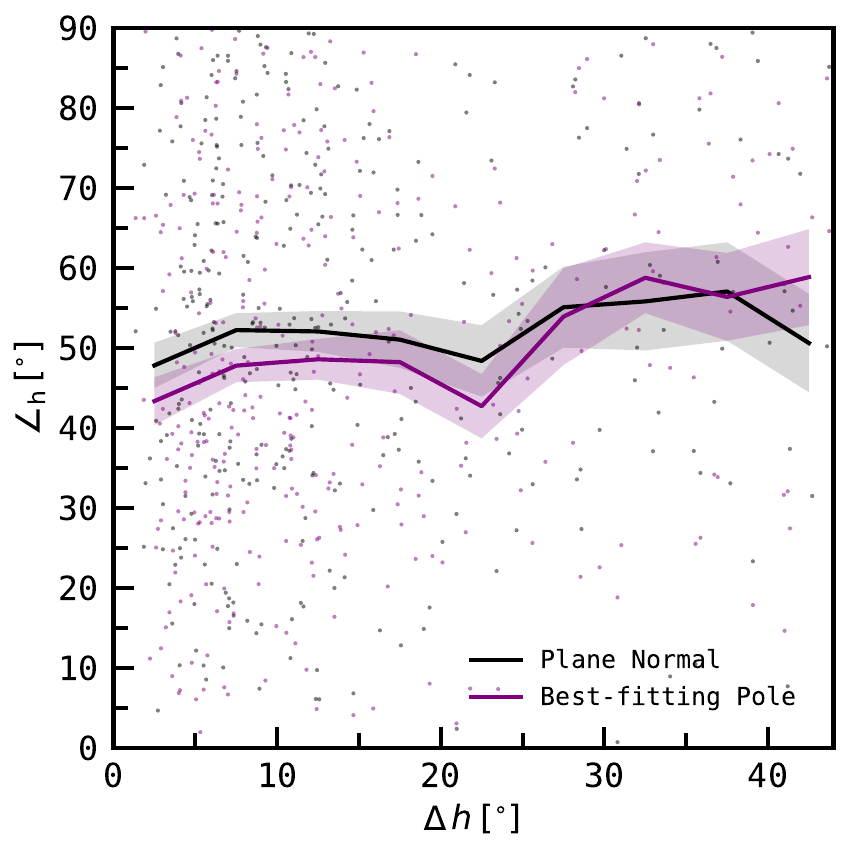}
	\includegraphics[width=0.311\textwidth]{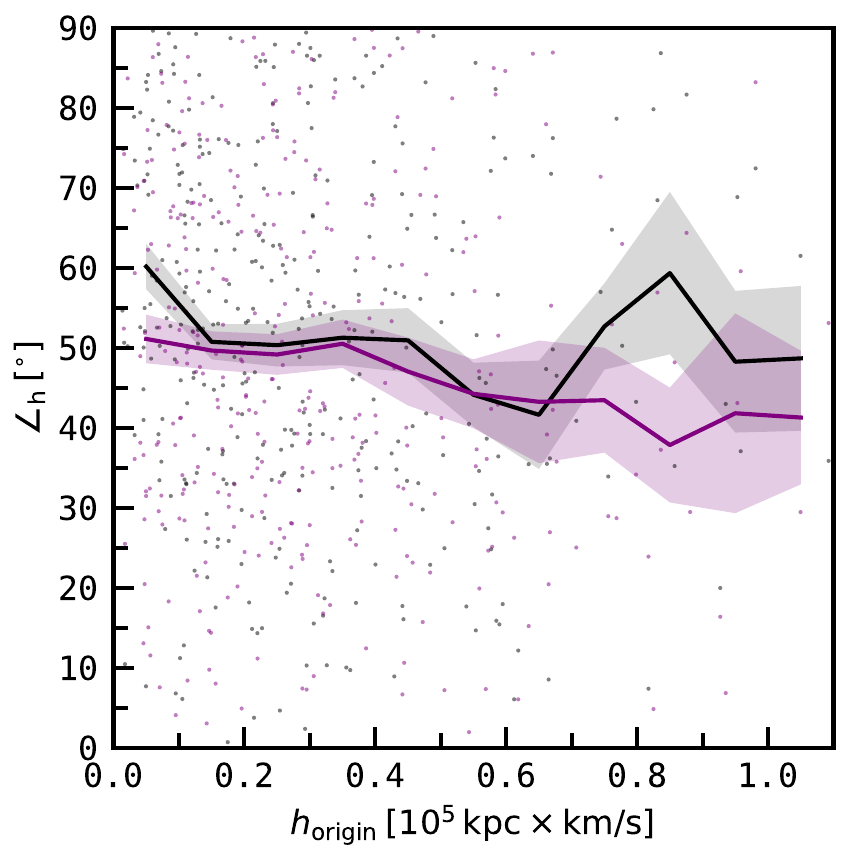}
    \caption{Interrelationships between merger specific angular momentum $h_{\mathrm{origin}}$, interaction pole instability $\Delta h$, and alignment $\angle_h$ with the present-time plane normal or the best-fitting satellite orbital pole in the \emph{full} sample. Means over binned systems are indicated by solid lines, and the standard error of these means are shaded in the corresponding colour. In the central and right-hand panels, black lines reflect the plane alignment $\angle_{h, \mathrm{plane}}$, whereas purple lines represent the orbital pole alignment $\angle_{h, \mathrm{orbit}}$.}
    \label{fig:s3_traj_intercorr}
\end{figure*}

Flattened satellite distributions formed by major mergers are expected to be roughly aligned with the interaction plane in both their spatial orientation and kinematics. We accordingly consider the angle between a merger's interaction pole and the plane normal ($\angle_{h, \mathrm{plane}}$) or best-fitting satellite orbital pole ($\angle_{h, \mathrm{orbit}}$) at present time. This merger alignment is plotted alongside specific angular momentum $h_{\mathrm{origin}}$ and interaction plane instability $\Delta h$ in Fig.~\ref{fig:s3_traj_dists}. Compared to an expected fraction of $1-\cos{30^{\circ}}=13.3$ per cent for alignments between fully isotropic vectors, we report that around 20 per cent of all \emph{merger-type} systems are aligned with either the plane normal or best-fitting pole to within $30^{\circ}$ -- thus demonstrating a $\sim50$ per cent enhancement over isotropy. While possibly reflecting a minority of mergers successfully forming a correlated present-time satellite distribution along its interaction plane, this result may also be a direct consequence of the orientation of local filaments, which define similar preferential directions for satellite infall and the approach of secondary merging haloes.

In the left and central panels in Fig.~\ref{fig:s3_traj_dists}, distributions are similar between \emph{full} and \emph{virial} samples for both $h_{\mathrm{origin}}$ and $\Delta h$ -- an expected result, since the dynamics of the host galaxies' merger is generally unrelated to the behaviour of their respective satellites. In the right-hand panel, the distribution of alignments $\angle_{h, \mathrm{plane}}$ and $\angle_{h, \mathrm{orbit}}$ are similar. $\angle_{h, \mathrm{orbit}}$ for \emph{full}-sample systems is noticeably lower than its \emph{virial} counterpart, while the same disparity is not seen for $\angle_{h, \mathrm{plane}}$. This kinematic signal may be driven by satellites slung out beyond their host's virial radius over the course of a merger, which would be disregarded when limiting our selection to the virial volume. In this case, the lack of a similar disparity for $\angle_{h, \mathrm{plane}}$ could imply that any slinging out of satellites caused by their hosts' merger only produces a degree of spatial anisotropy comparable to or weaker than natural anisotropies resulting from preferential directions of satellite infall. While the result could also be due to the present-time system's halo shape aligning with the aforementioned preferred directions -- an effect which may be less-pronounced in the central, $<R_{200}$ region -- this mechanism should also enhance $\angle_{h, \mathrm{plane}}$ in the \emph{full} sample, an effect not seen in Fig.~\ref{fig:s3_traj_dists}.

\begin{figure*}
	\includegraphics[width=0.32\textwidth]{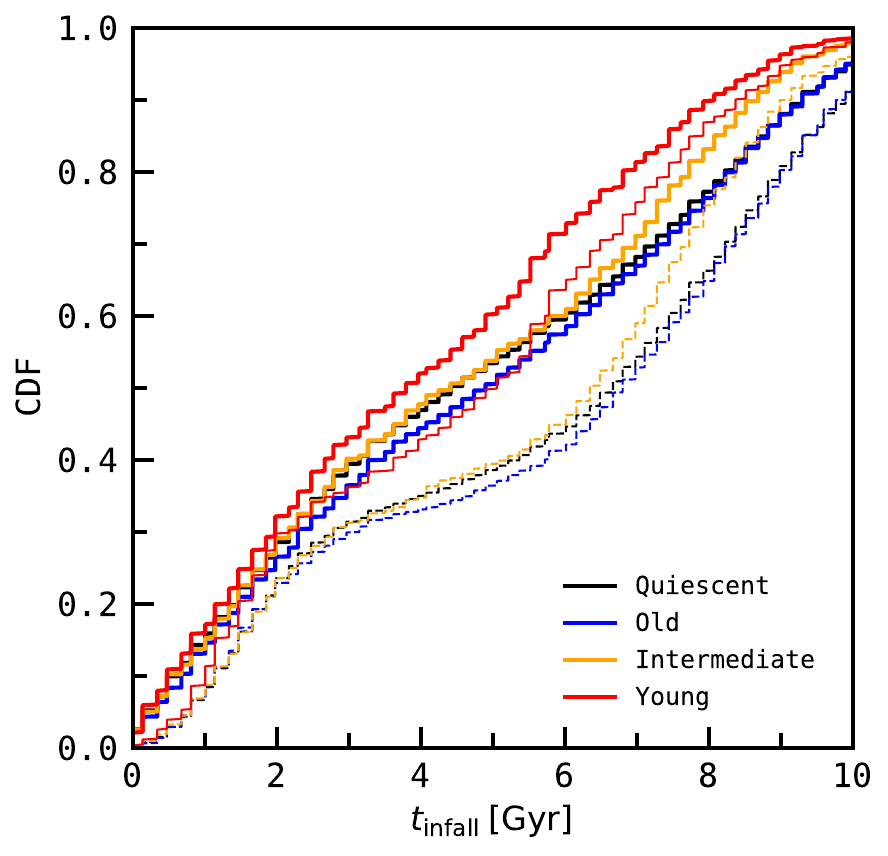}
	\includegraphics[width=0.316\textwidth]{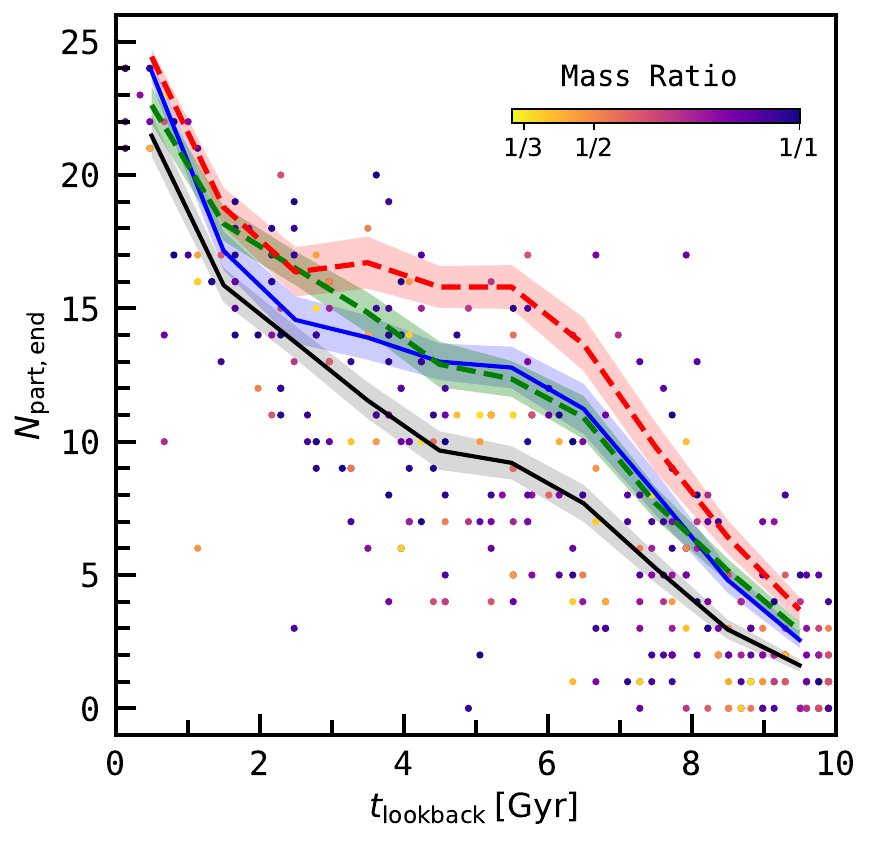}
	\includegraphics[width=0.316\textwidth]{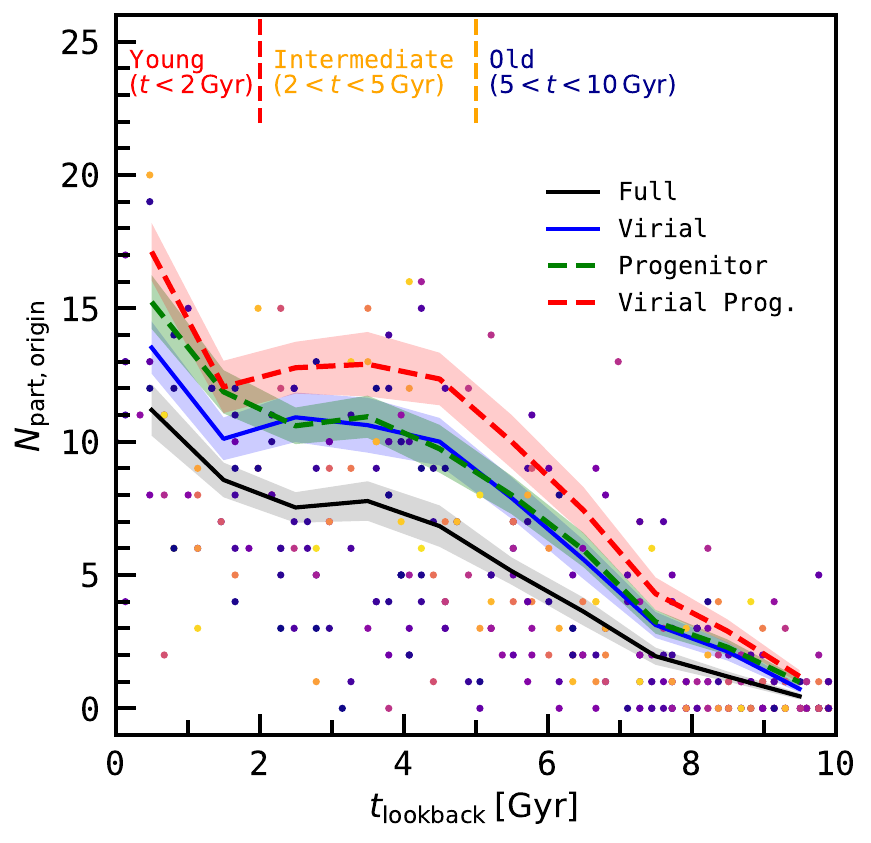}
    \caption{Accretion times of simulated satellites, defined as the first snapshot where the corresponding subhaloes are considered to be bound to their current host by \texttt{Subfind}. \textbf{Left Panel:} Infall lookback times for our present-time sample of satellite subhaloes, categorised by their host galaxy's merger history. Satellites from the \emph{full} and \emph{virial} samples are indicated by thick and thin lines respectively. Refer to Table~\ref{tab:s3_categories} for system category definitions. \textbf{Central Panel:} The number of satellites per system bound to their host halo by the end of its last merger, plotted against lookback time to the merger's completion. Each data point represents a simulated \emph{full}-sample system and is coloured by its merger's mass ratio. Means over binned systems are indicated by solid black and blue lines -- corresponding to \emph{full} and \emph{virial} systems respectively -- and the standard error of these means are shaded in the corresponding colour. We additionally plot binned means and their standard error for \emph{progenitor} and \emph{virial progenitor} satellites ranked by their maximum past mass in green and red respectively. \textbf{Right Panel:} Same as the central panel, but instead shows the number of satellites bound to their host by the \emph{origin} of of their last merger. Lookback time thresholds between \emph{young}, \emph{intermediate}, and \emph{old} systems are indicated by red and orange dotted lines.} 
    \label{fig:s4_infall}
\end{figure*}

\begin{figure}
    \hspace{10mm}
	\includegraphics[width=0.35\textwidth]{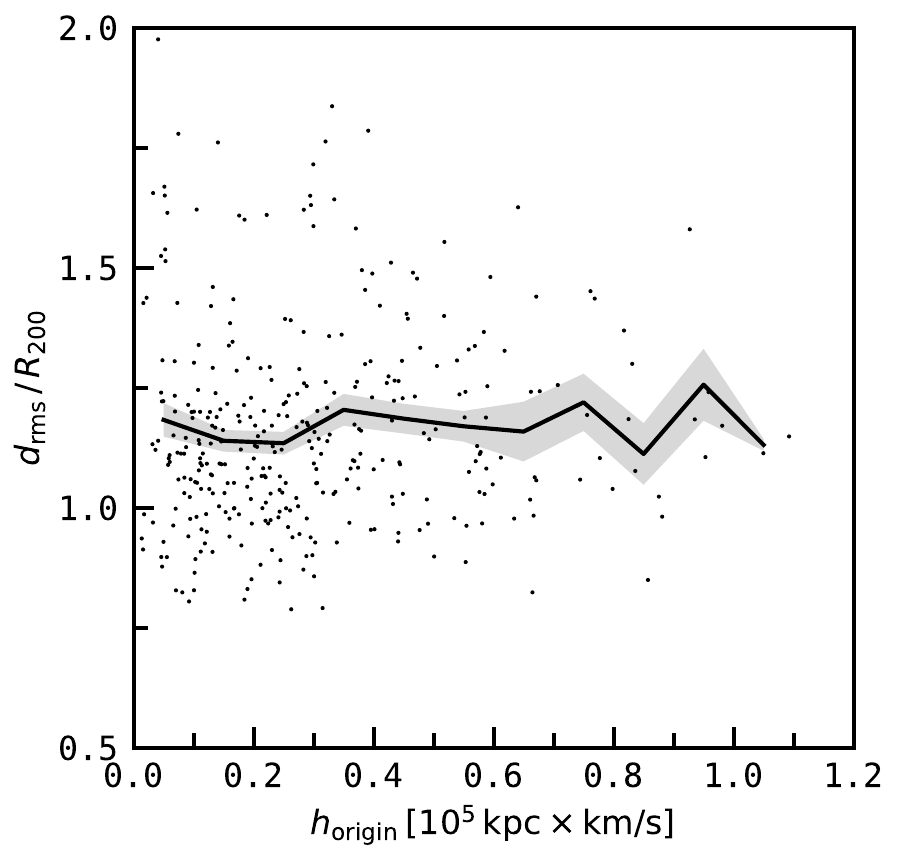}
    \caption{
        Present-time satellite radial extent $d_{\mathrm{rms}}/R_{200}$ plotted against the initial specific angular momentum of their host's last major merger $h_{\mathrm{origin}}$. Black dots present individual systems, while the black line denotes a running mean over equally spaced $h_{\mathrm{origin}}$ bins. The shaded region represents the corresponding standard error in the mean.
    }
    \label{fig:extent_vs_angmom}
\end{figure}

We now compare the degree of phase-space correlation found in well-aligned ($\angle_{h} < 30^{\circ}$) and less-aligned ($\angle_{h} \geq 30^{\circ}$) systems in Fig.~\ref{fig:s3_traj_kde}. \emph{Satellite planes well-aligned with their last merger's interaction plane} \emph{tend to be more flattened} (with median shifts in $c/a$ and $\Delta_{\mathrm{rms}}$ of 0.07 and 12 kpc respectively), whereas a similar improvement is not seen for well-aligned best-fitting poles (except for $c/a$, which shows a markedly reduced shift of 0.03). Indeed, we report a weak but significant positive correlation between $c/a$ and $\angle_{\mathrm{h, plane}}$ (Kendall: $p=2.2\times10^{-3}$) while no associated correlation with $\Delta_{\mathrm{rms}}$ is found, a result similar to what one would expect in \citetalias{Smith2016merger}'s model wherein the distribution of satellites along the merger's interaction pole is roughly preserved. However, this enhancement in satellite correlation is not necessarily due to the slinging out of satellites -- radial extents for well-aligned and less-aligned systems are statistically indistinguishable in $\angle_{h, \mathrm{plane}}$ (KS: $p=0.75$).
We additionally find that mergers with higher $h_{\mathrm{origin}}$ result in thicker plane heights at present time (Kendall: $p=1.8\times10^{-6}$). No corresponding significant correlation exists in the scale-free $c/a$, however, suggesting a systematic effect -- systems with larger characteristic distances and velocities inherently have higher specific angular momenta.

Finally, we search for interrelationships between our merger trajectory parameters in Fig.~\ref{fig:s3_traj_intercorr}. 
In the left-hand panel, high-angular momentum mergers correlates with a lower degree of interaction plane instability (Kendall: $p=2.2\times10^{-4}$), which in turn weakly correlate with better interaction pole alignments with best-fitting orbital poles at present time (Kendall: $p=6.1\times10^{-3}$) -- the corresponding correlation with plane normal alignments does not appear to be statistically significant.
If mergers efficiently sling out satellites along their interaction plane, we would expect to find many systems with low $\angle_{h}$, especially with correspondingly large $h_\mathrm{origin}$. And yet, $h_\mathrm{origin}$'s negative correlation with interaction pole alignment is marginal at best (Kendall: $p_{\mathrm{plane}}=1.1\times10^{-2}$, $p_{\mathrm{orbit}}=3.4\times10^{-3}$). We do note that \emph{virial} systems only demonstrate a negligible correlation with either alignment mode (Kendall: $p_{\mathrm{plane}}=0.29$, $p_{\mathrm{orbit}}=0.25$) which may hint at a minority of satellites successfully having been slung out past their halo's virial volume, but their contribution is inconclusive.

Overall, we argue that mergers -- even those with high specific angular momenta -- appear to be highly inefficient at best in forming correlated satellite distributions as per \citetalias{Smith2016merger}'s model. Indeed, mergers with high specific angular momenta do not significantly produce more radially extended satellite distributions than those which follow near-radial trajectories (Kendall: $p=0.09$, see Fig.~\ref{fig:extent_vs_angmom}).

\begin{figure*}
	\includegraphics[width=0.333\textwidth]{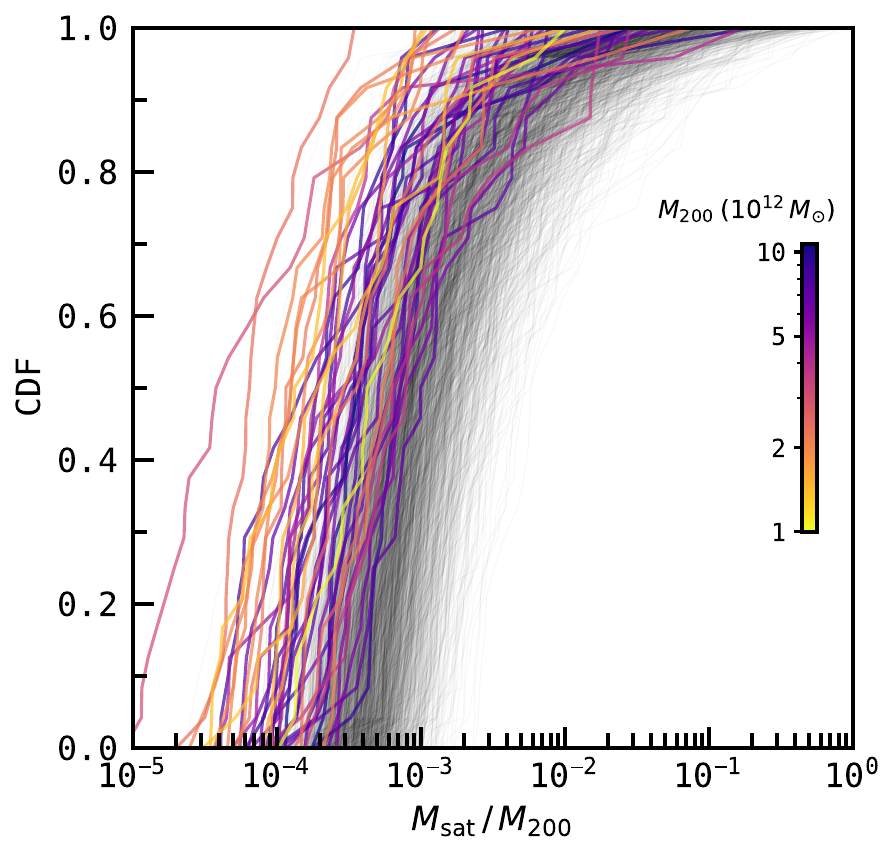}
  \includegraphics[width=0.32\textwidth]{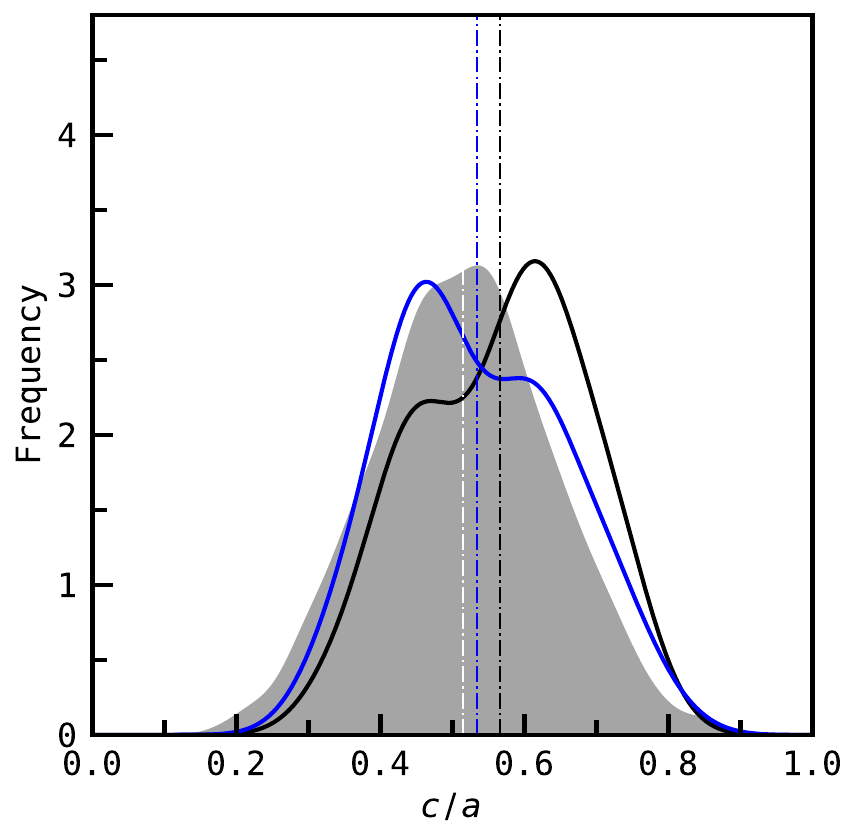}
	\includegraphics[width=0.309\textwidth]{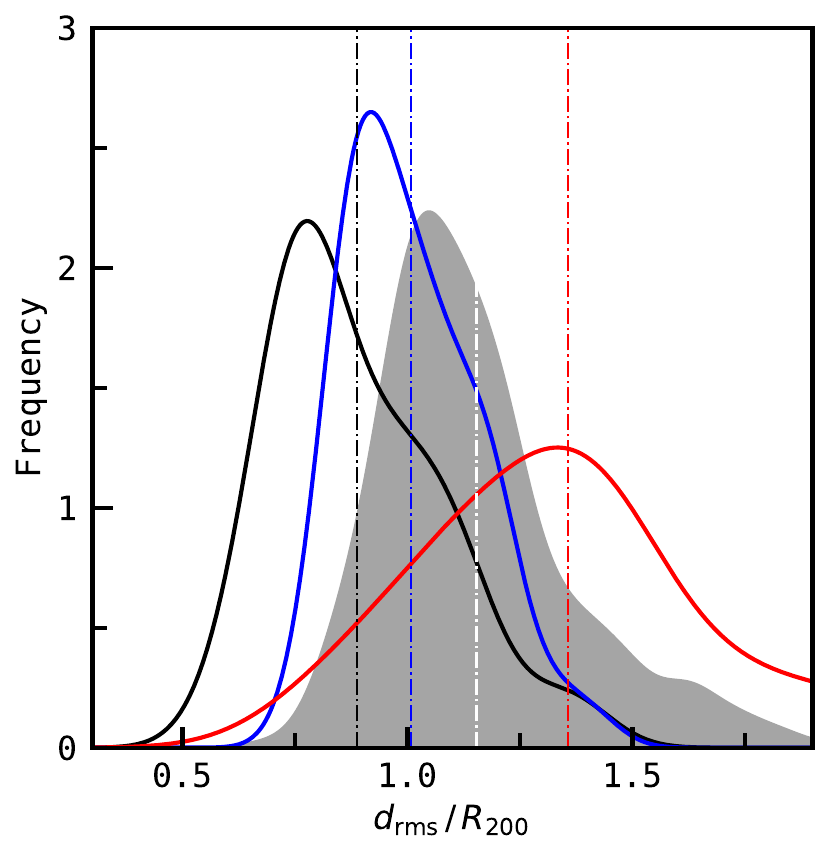}
	\includegraphics[width=0.325\textwidth]{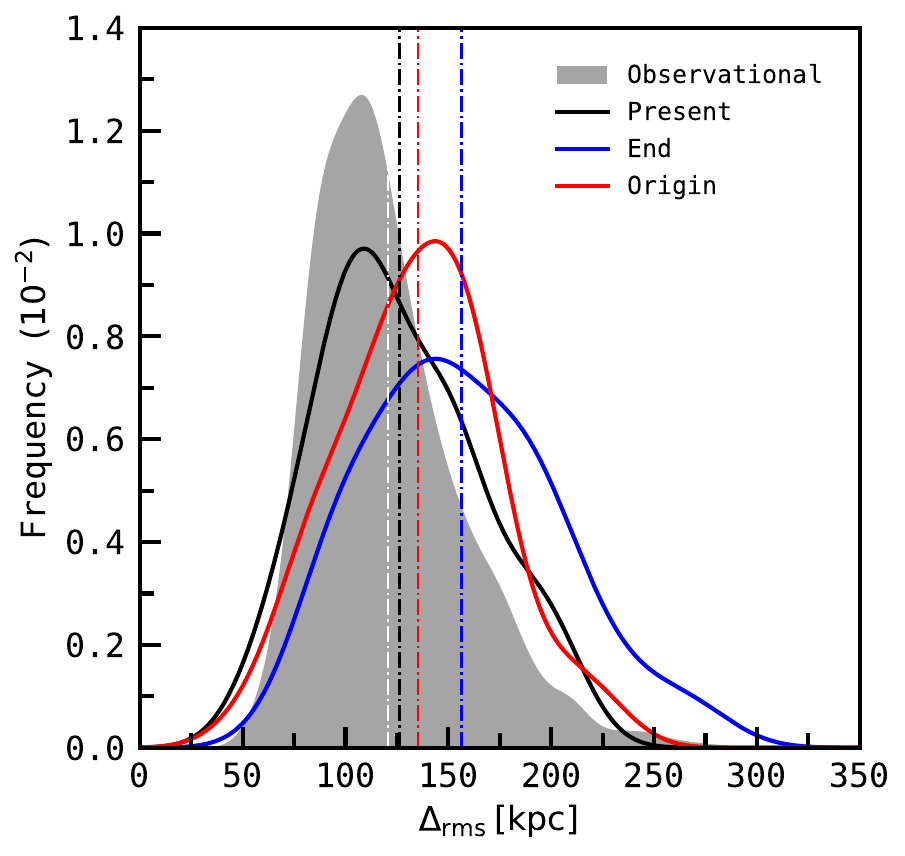}
	\includegraphics[width=0.322\textwidth]{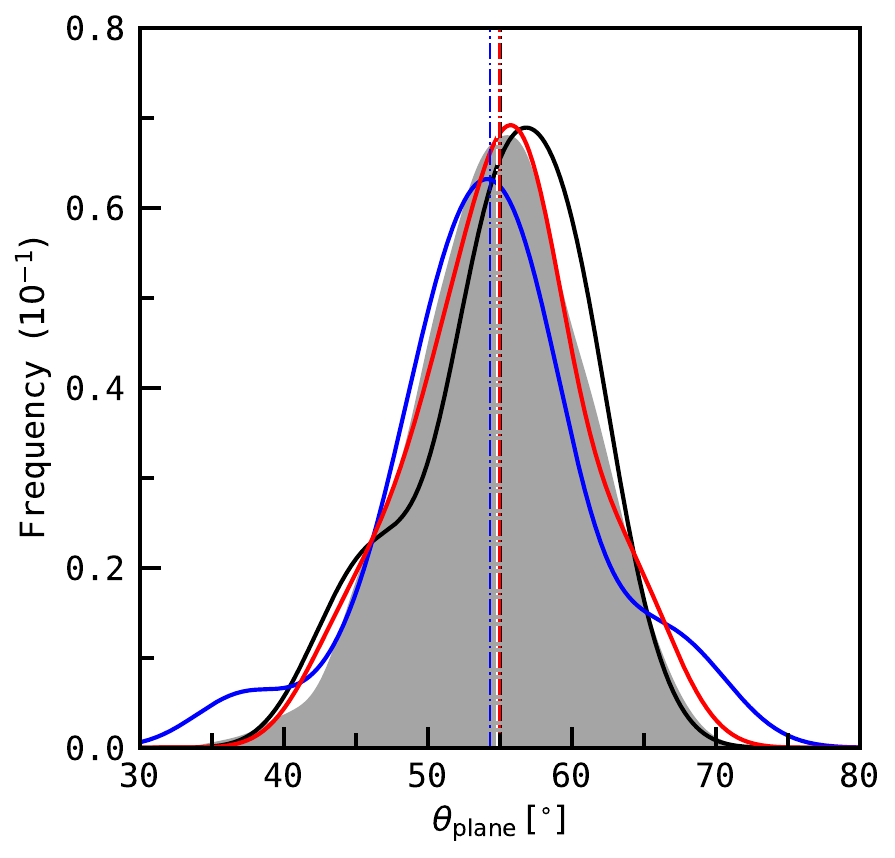}
	\includegraphics[width=0.318\textwidth]{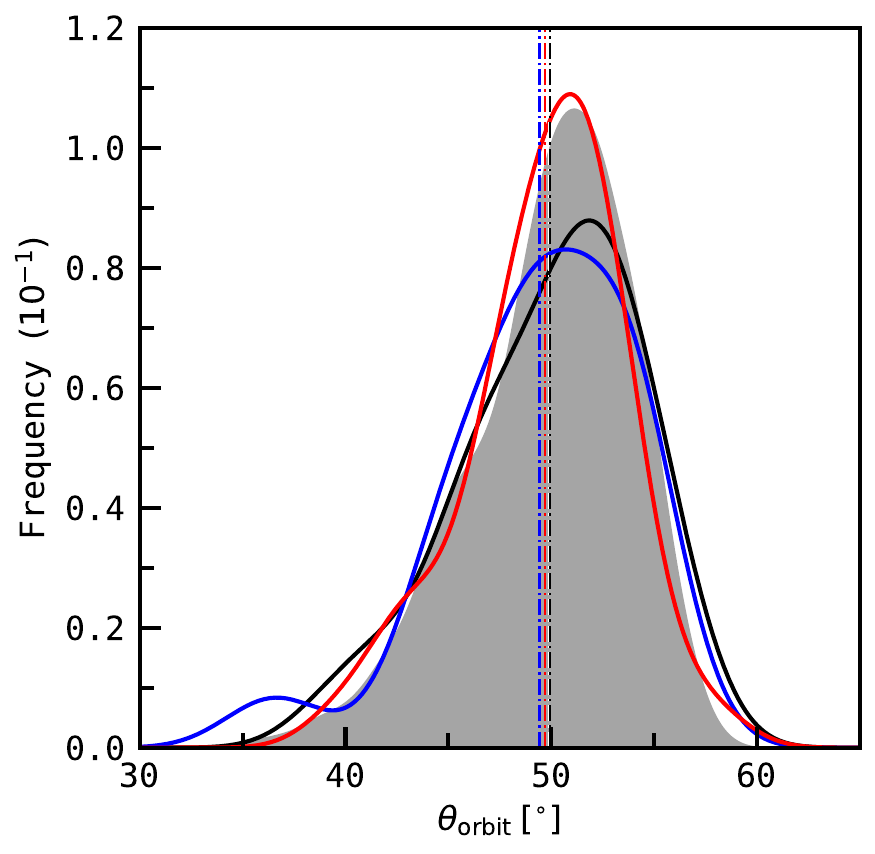}
    \caption{\textbf{Upper Left Panel:} Cumulative distributions of satellite masses per system, normalised to their host halo's virial mass. Black lines represent observationally motivated samples of the 25 most massive satellites at present time. Distributions for the 25 most massive satellites that were bound to one of the two merging haloes at their merger's \emph{origin} -- referred to as 'participant' satellites -- are coloured by their present-time host halo's virial mass, and are themselves generally around an order of magnitude less massive than the observationally motivated sample (a trend also seen when comparing stellar masses). \textbf{Remaining Panels:} Phase-space correlation metrics of participant satellite distributions at the \emph{origin} and \emph{end} of their system's last major merger, as well as at present time. The present-time metric distribution for observationally motivated systems is also shown in shaded grey. Means of each distribution are indicated by dashed lines in white for the observationally motivated sample, and the corresponding colour for participant satellites.} 
    \label{fig:s4_kde}
\end{figure*}

\section{Satellite Participation}
\label{sec:s4}

\subsection{Infall times and merger participation}
\label{sec:s4_infall}

Thus far, we have been unable to identify a substantial imprint of major mergers on present-time satellite distributions. Here, we argue that such an imprint may be partially washed out by the post-merger accretion of satellites, resulting in a significant fraction of the most massive satellites at present time not having participated in the system's last merger. To quantify this bias, we define a satellite's (lookback) infall time $t_{\mathrm{infall}}$ as the first snapshot where it is recognized as bound to its current host halo by \texttt{Subfind}. If a satellite is accreted by a halo before being temporarily flung out, we record the time for the initial accretion event.

Infall times for all systems (including those without a major merger on record) are shown in Fig.~\ref{fig:s4_infall}. There appears to be a slight bimodality in $t_{\mathrm{infall}}$, with peaks around 3 and 6 Gyr ago. Satellites in the \emph{virial} sample generally appear to be accreted earlier than in the \emph{full} sample, a difference most likely due to the former's central radial distribution -- the \emph{virial} satellites' smaller host-centric distances reflect a longer period of time under the effects of dynamical friction, corresponding to earlier infall times. In addition, \emph{virial} satellites would experience earlier accretion events when their host was less massive -- corresponding to lower-energy orbits that remain within the virial volume today. Satellites in \emph{young} systems tend to have been accreted more recently, but only show a significant difference with cumulative distributions for \emph{old} and \emph{quiescent} satellites with lookback times of 3 Gyr or larger -- far before their last major mergers ended (within the last 2 Gyr). As only satellite accretion events onto main progenitors of the present-time halo are considered here, this trend implies that satellites from the less-massive merging system are captured by the primary halo while their merger is ongoing (the duration of which, from our definition in Section~\ref{sec:s3_identify}, generally ranges from $1.5-5\,\mathrm{Gyr}$).

The central and right-hand panels in Fig.~\ref{fig:s4_infall} compare the lookback time to the mergers' completion with the number of 'participant' satellites per system, which we define as being bound to one of the two merging haloes before the end ($N_{\mathrm{part, end}}$) or the beginning ($N_{\mathrm{part, origin}}$) of the system's last major merger. As expected, an earlier merger corresponds to a smaller number of participants. For a merger event that ended 5 Gyr ago, up to $20/25$ satellites would have been accreted after its completion (right panel in Fig.~\ref{fig:s4_infall}). Very few of the most luminous present-day satellites experienced the whole duration of a merger event, with $N_{\mathrm{part, origin}}$ generally only reaching $60$ per cent of the satellite population, even for very recent mergers ($t_{\mathrm{lookback}} < 1\,\mathrm{Gyr}$). We argue the true number of satellites strongly influenced by a merger's angular momentum imprint should lie between $N_{\mathrm{part, end}}$ and $N_{\mathrm{part, origin}}$, which respectively serve as upper and lower extremes. A merger's mass ratio does not appear to have a robust impact on the participation of present-time satellites when holding $t_{\mathrm{lookback}}$ constant. 

Satellites in the \emph{virial} sample demonstrate a higher mean $N_{\mathrm{part}}$ than those drawn from the \emph{full} sample, which is likely a direct result of the former's earlier infall times. And yet, spatial anisotropies in IllustrisTNG's satellite distributions appear to be driven by satellites beyond their halo's virial radius (see Section~\ref{sec:s2_psc}), which are -- by definition -- disregarded in our \emph{virial} sample. This highlights the difficulty in retaining enough satellites with a strong merger imprint that also lie at large host-centric distances, and suggests the impact of merger events on satellite distributions may be short-lived in a cosmological context.

\begin{figure*}
	\includegraphics[width=0.315\textwidth]{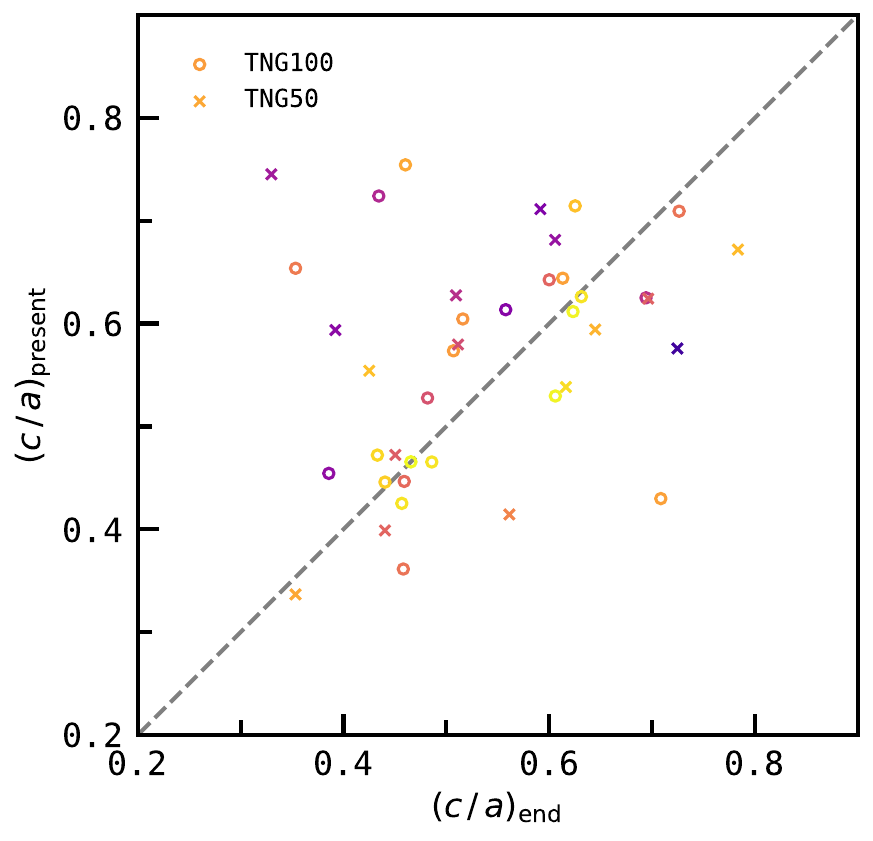}
	\includegraphics[width=0.316\textwidth]{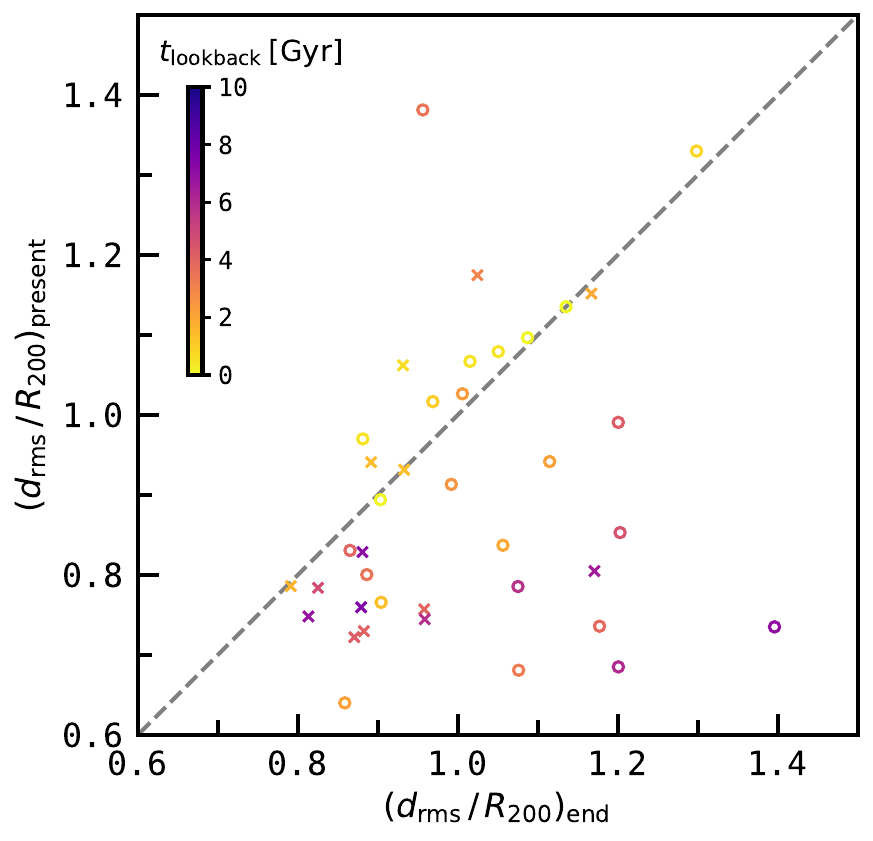}
	\includegraphics[width=0.322\textwidth]{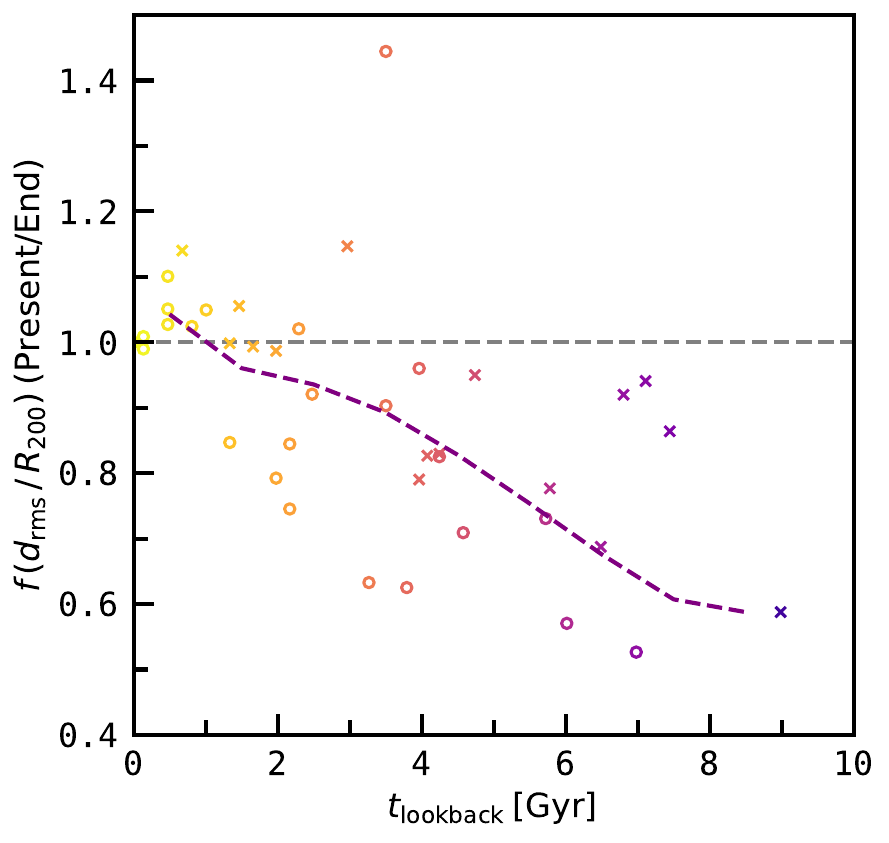}
    \caption{\textbf{Left Panel:} Post-merger changes in the minor-to-major axis ratio $c/a$ of participant satellite distributions. Circles and crosses represent individual systems sampled from TNG100-1 and TNG50-1 respectively, and are coloured by the time since their last major merger ended, $t_{\mathrm{lookback}}$. The grey dashed line divides systems that experience a net gain (upper left) or net loss (lower right) in $c/a$ between $t_{\mathrm{lookback}}$ and present time. \textbf{Central Panel:} Same as the left-hand panel, but for satellite radial extent $d_{\mathrm{rms}}/R_{200}$. \textbf{Right Panel:} The change in $d_{\mathrm{rms}}/R_{200}$ between the merger's \emph{end} and present time for participant satellites, represented as a fraction -- a decrease in radial extent corresponds to $f<1$. The binned mean for the concatenated sample of TNG100-1 and TNG50-1 systems is drawn by the purple line.}
    \label{fig:s4_change}
\end{figure*}

When ranking satellites by their peak mass instead, we find a consistent mean increase of around $2-3$ satellites in $N_{\mathrm{part}}$ when holding the adopted satellite search radius constant. Satellites in the \emph{progenitor} and \emph{virial progenitor} samples are expected to have experienced mass stripping from dynamical friction for a longer period than \emph{full} and \emph{virial}-sample counterparts, and hence display earlier accretion times than the most massive satellites at present time. However, many of the traits founds in the \emph{full} and \emph{virial} sample distributions can also be seen here. A sharp decline in $N_{\mathrm{part}}$ between $t_{\mathrm{lookback}} = 0-2\,\mathrm{Gyr}$ is followed by a shallow, plateau-like region until $5-6\,\mathrm{Gyr}$ -- the latter appears to be most prominent in the \emph{virial progenitor} sample's mean. Assuming satellites with high peak masses retain most of their initial stellar content and are sufficiently luminous at present time, it may be possible to detect signatures for major mergers as old as $6\,\mathrm{Gyr}$, only washed out by as little as $\sim25$ per cent of the current satellite population from post-merger accretion events.

We also briefly check whether the phase-space correlation of present-time satellite distributions depends on their system's participant populations $N_{\mathrm{part, end}}$ and $N_{\mathrm{part, origin}}$, but recover no statistically significant result. If mergers indeed leave an imprint on satellite systems that improve their degree of correlation, it is likely a small effect that is only seen for specific mergers rather than merger events in general, and is difficult to recognize even in a subsample of high-$N_{\mathrm{part}}$ systems. The post-merger accretion of new subhaloes only exacerbates this problem.

\subsection{Participant satellite distributions}
\label{sec:s4_participants}

To isolate the imprint of major mergers from the washing-out effect arising from any post-merger satellite accretion, we now find systems purely consisting of participant satellites accreted before their last merger's \emph{origin} -- ensuring they would have experienced the full duration of the merger and maximising any imprint left. For each \emph{merger}-type system (with a major merger within the last 10 Gyr), we identify all subhaloes bound to their host halo at present time, and rank them by total mass. For each satellite, we check whether it was bound to one of the two merging haloes at the last merger's \emph{origin}, stopping once 25 participants are found, 1000 satellites are checked or no satellites remain. Since small satellite populations systematically appear more correlated, all systems with less than 25 participants are rejected, leaving only 22 valid systems. To improve the size of our participant sample, we also repeat the above procedure in the higher-resolution TNG50 run, gaining 16 systems for a total of 38.

We define this participant satellite sample alongside the previously adopted sample of the most massive satellites at present time. We henceforth refer to the latter as the "observationally motivated" sample, since it is roughly analogous to the sample which would be observed if the simulated systems were real. On the other hand, the participant satellites have experienced tidal mass stripping for a comparatively longer duration due to their earlier infall compared to the most massive satellites at present time. As a result, their total mass distribution shown in the upper left-hand panel of Fig.~\ref{fig:s4_kde} is around an order of magnitude less massive than observationally motivated satellites -- we confirm the same trend holds for their stellar mass content.

We find distributions of participant satellites and calculate their associated phase-space correlation metrics for their last merger's \emph{origin} and \emph{end}, as well as at present time. Participant distributions are shown for our phase-space correlation metrics in Fig.~\ref{fig:s4_kde}, along with those of their observationally motivated counterparts for reference.

We note that phase-space correlation metrics for participant satellites at their merger's \emph{origin} are systematically biased towards highly flattened and extended distributions, as satellites are then distributed around both merging host galaxies. The major axis of the combined satellite distribution is dominantly defined by the vector connecting one host to another. Since the ToI fit used to derive the system's $c/a$ axis ratio is sensitive to outliers \citep{Pawlowski2015problems}, even mergers with a comparatively small mass ratio -- with a correspondingly small fraction of the satellite population initially bound to the secondary halo -- return a skewed $c/a$ distribution. As such, we do not include the $c/a$ distribution at the merger \emph{origin} in the upper central panel in Fig.~\ref{fig:s4_kde}.

When comparing the present-time distributions of observationally motivated and participant satellites, the latter demonstrates a slightly higher mean $c/a$ by 0.05. A two-sample KS test reveals that the two distributions are distinguishable ($p=7\times10^{-3}$), although the significance of this shift is unreliable due to the small sample size. Visually, we find a notable gain in less-flattened ($c/a>0.55$) participant systems with respect to the observationally motivated sample.

Conversely, there is a marked difference between the radial extents of observationally motivated and participant satellite distributions. Participant satellites with a full merger imprint already form compact distributions ($d_{\mathrm{rms}}/R_{200}=1.01\pm0.14$) at the end of their hosts' merger. At present time, participant distributions are increasingly compact, with a mean shift of 0.27 compared to $d_{\mathrm{rms}}/R_{200}=1.15\pm0.22$ from the observationally motivated sample (KS: $p=10^{-12}$).
Meanwhile, we find that the absolute plane heights of participant systems are largest at their merger's \emph{end}, eventually decreasing to a present-time distribution of $\Delta_{\mathrm{rms}}=126\pm38\,\mathrm{kpc}$ fully consistent with plane heights of our observationally motivated sample. From the lack of a corresponding decrease in $c/a$ over this period, we presume this shift is driven mainly by the diminishing spatial extent of the system. 
Kinematic metrics $\theta_{\mathrm{plane}}$ and $\theta_{\mathrm{orbit}}$ are statistically indistinguishable between observationally motivated and participant systems at all three key merger stages.

In Fig.~\ref{fig:s4_change}, we focus on the post-merger shift in spatial correlation metrics for individual systems. A given system of participant satellites has a similar likelihood of gaining or losing $c/a$ after their host's merger reaches completion until present time. For the left-hand panel, we find a Kendall coefficient of $\tau=0.23$ ($p=0.036$) -- while some correlation between a system's $c/a$ at $t_{\mathrm{lookback}}$ and present time may exist, the weakness of the obtained correlation suggests that the flattening of these systems may mostly be a result of spurious, short-lived satellite alignments.

Conversely, only 11 out of 38 systems experience a net gain in $d_{\mathrm{rms}}/R_{200}$ over this period -- of these, only 3 demonstrate a $d_{\mathrm{rms}}/R_{200}$ growth of over $10$ per cent, whereas many systems with reduced radial extent show a $20-30$ per cent decrease. Overall, a majority of participant distributions grow more compact after their merger ends. The right-hand panel further shows a distinct negative correlation between each system's post-merger fractional change in radial extent and the lookback time to its last major merger's completion (Kendall: $\tau=-0.49$, $p=6.9\times10^{-6}$) -- this shrinking in satellite extent is continuous even after their host's merger ends.

Since the participant satellite systems' $d_{\mathrm{rms}}/R_{200}$ distribution on merger completion is already distinctly lower than the that for observationally motivated systems at present time, we interpret this post-merger contraction of radial extent as a consequence of tidal disruption from the merger itself in addition to continuous energy loss from dynamical friction. Flattening is, on average, conserved throughout this process, although whether a given system gains or loses $c/a$ after its last merger appears stochastic. Kinematic correlation is persistently weak in both participant and observationally motivated satellites, and does not change after experiencing a host merger. \emph{Overall, we find no indication of mergers enhancing the phase-space correlation of participant systems when compared to the observationally motivated sample of the most massive satellites at present time.}

\section{Conclusions}
\label{sec:s5}

\citet{Smith2016merger} (\citetalias{Smith2016merger}) previously reported that high-mass ratio galaxy mergers in an $N$-body simulation with near-circular infall trajectories can imprint their angular momenta on their satellite galaxy populations -- slinging individual satellites outward to large radii and consequently forming thin, co-rotating and long-lived planes-of-satellites. 
Motivated by their results, we searched for the imprint of major mergers on present-time satellite galaxy distributions in the IllustrisTNG suite of hydrodynamic cosmological simulations.

In general, we find that major mergers with mass ratios of $1/3$ and higher have a statistically negligible impact on the phase-space correlation of the most massive satellites at present time, with hints of a marginally adverse effect on the satellites' spatial and kinematic coherence. Systems that have experienced a recent merger within the last 2 Gyr weakly tend towards more-isotropic satellite distributions with larger minor-to-major axis ratios. 
We found little evidence of the proposed slinging out of satellites -- mergers with less-radial infall trajectories (traced by a high specific angular momentum content) do not produce any more radially extended satellite distributions nor enhance the phase-space correlation of present-time satellites than mergers on semi-radial trajectories.

A higher angular momentum does correlate with an improved stability in the interaction plane along which the merger occurs. A higher stability, in turn, slightly enhances the alignment between the merger interaction axis and the minor axis of the present-time satellite distribution, implying a more successful imprint of the merger's angular momentum onto the participating satellite distributions. The frequency of such well-aligned systems in our simulated sample is around $50$ per cent higher ($20$ against $13$ per cent) than expected in a fully isotropic satellite distribution.

Nevertheless, well-aligned systems marginally enhance their satellite systems' degree of spatial correlation, though the improved flattening is driven by a lower absolute plane height rather than a more extended distribution. This weak trend does point in the expected direction (wherein mergers with the right conditions can strengthen satellite phase-space correlation) but is by far not a dominant effect -- in most simulated systems, the disruptive impact of their last major mergers appears to prevail.

The absence of evidence for major mergers slinging out satellite galaxies may be due in part to the radial distribution of subhaloes in TNG, which extend past their host halo's virial radius up to several $R_{200}$. The simulated distributions are similar to that of the Centaurus A satellites and consistent with the M31 system, but are far more extended than both the classical and fainter Milky Way satellites. Conversely, \citetalias{Smith2016merger} generates a highly constrained satellite distribution initially localised within $<0.25\,R_{200}$ of a Milky Way-mass halo, resulting in the efficient slinging out of satellites to radii more comparable to systems in a full cosmological context.

In our sample, we would expect that satellites initially closer to their host halo's centre, if any, are slung out effectively to form a coherent plane. However, the spatial anisotropy in simulated systems appears to be primarily driven by the outermost satellites. As found in Section~\ref{sec:s2_volume}, satellite systems confined to the virial volume of most host galaxies are more isotropic than the distribution of more distant satellites, making it unlikely to find an initially thin distribution (as required by \citetalias{Smith2016merger}'s conditions) that would also be sufficiently centrally concentrated to be susceptible to being slung out.

Furthermore, the post-merger accretion of fresh satellites has a significant contribution to the sample of the most massive satellites at present time, thus washing out any potential merger imprint. Systems with major mergers around 5 Gyr ago tend to have accreted more than 50 per cent of their present-time satellite population post-merger. In addition, satellites bound to the merging haloes from the merger's beginning are rare -- even systems with recent mergers within the last $1-2\,\mathrm{Gyr}$ are not populated with more than 60 per cent of such satellites. When considering a separate sample of such 'participant' satellites that experienced the full merger event, we find no particular improvement in phase-space correlation. In addition, the stellar masses of participant satellites at present time are approximately an order of magnitude lower than the most massive satellites, and thus are unlikely to form a majority in an observational sample. Finally, participants experience a continuous shrinking in their radial distribution throughout and after their system's last merger due to the effects of dynamical friction and disruptions from said merger, while -- on average -- their flattening and kinematic correlation is preserved.

As a caveat, it should be noted that \citetalias{Smith2016merger}'s model of satellite plane formation via mergers was modelled in early-universe, $z=3$ ($t_{\mathrm{lookback}}=11.6\,\mathrm{Gyr}$) haloes, with a virial mass comparable to the present-day Milky Way and M31 but a radius around a quarter of that expected at present time -- resulting in highly concentrated satellite distributions. Since haloes accrete mass throughout its cosmic lifetime via hierarchical clustering in the concordance $\Lambda$CDM model of structure formation, we would expect descendants of \citetalias{Smith2016merger}'s haloes to be far more massive than the Local Group hosts at present time. The success of \citetalias{Smith2016merger}'s model is likely due to this initial concentration of satellites, which in turn enables mergers to efficiently sling satellites outward to relatively large radii (at least, compared to the initial satellite radial distribution). However, this now creates a \emph{catch-22} scenario -- the formation of satellite planes via mergers is more efficient at higher redshift, but the newly formed planes are unlikely to survive until present time due to washing-out from post-merger satellite accretion. On the other hand, recent major mergers do not frequently form correlated planes of satellites. Consequently, we argue that mergers of host galaxies are not an effective, general solution to the apparent ubiquity of satellite planes in the local Universe.

In rare cases where mergers efficiently transfer their angular momentum to the present-time satellite distribution, the resulting system can demonstrate a marginally higher degree of phase-space correlation. If the merger is recent ($<2\,\mathrm{Gyr}$), the comparatively low number of satellites accreted post-merger may permit the observational confirmation of such a merger imprint. We emphasize that our results do not rule out an individual merger with favourable initial conditions forming a correlated distribution, but rather only show that such occurrences are rare in a cosmological context. It is still conceivable that the observed Centaurus A plane -- flattened not by a thin absolute plane height but from an extended radial distribution \citep{Muller2019casp} -- may have been formed from its proposed merger 2 Gyr ago \citep{Wang2020merger}.

\section*{Acknowledgements}

We thank the anonymous referee for their thoughtful and helpful comments. KJK and MSP acknowledge funding via a Leibniz-Junior Research Group (project number J94/2020). MSP also thanks the German Scholars Organization and Klaus Tschira Stiftung for support via a KT Boost Fund.

\section*{Data Availability}

The data underlying this article will be shared on reasonable request to the corresponding author. Data products in the IllustrisTNG suite are publicly available at \texttt{www.tng-project.org}.



\bibliographystyle{mnras}
\bibliography{paper} 




\appendix

\section{Impact of system scale}
\label{sec:aa}

\begin{figure}
	\includegraphics[width=0.4\textwidth]{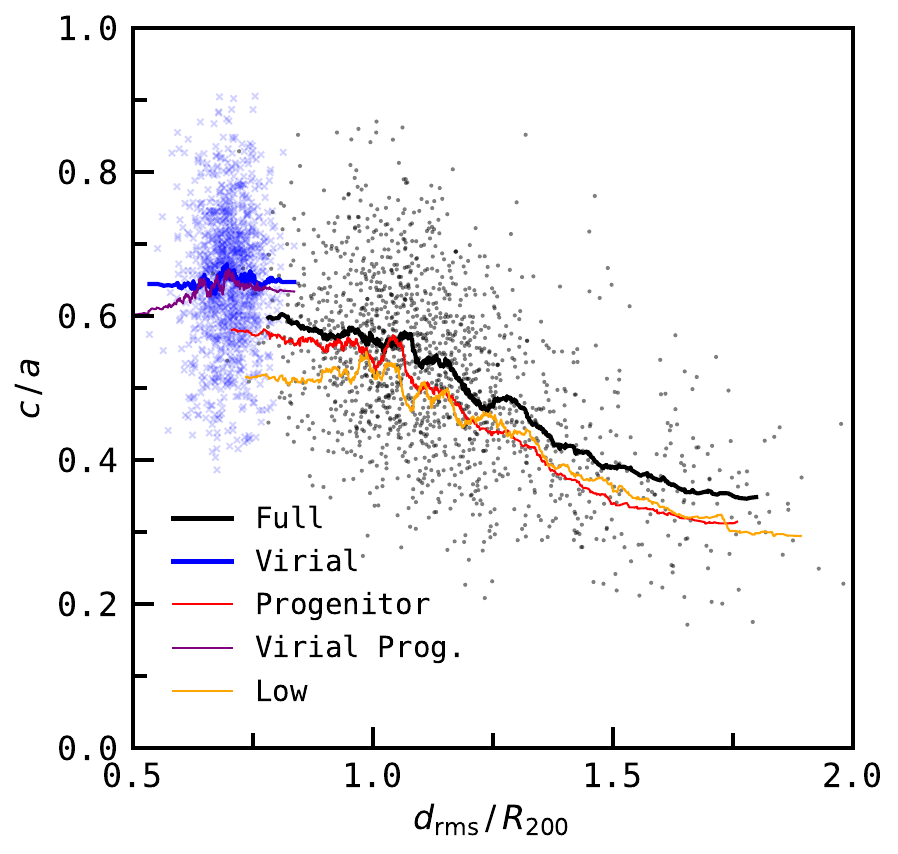}
    \caption{
        Best-fitted plane flattening over a range of radial extents. Black points and blue crosses represent individual simulated systems in the \emph{full} and \emph{virial} samples respectively, while solid lines in the corresponding colours represent their rolling average. Rolling averages for systems in the \emph{progenitor}, \emph{virial progenitor}, and \emph{low} samples are overlaid as red, blue, and orange lines respectively.
    }
    \label{fig:aa_extent_ca}
\end{figure}

\begin{figure*}
	\includegraphics[width=0.394\textwidth]{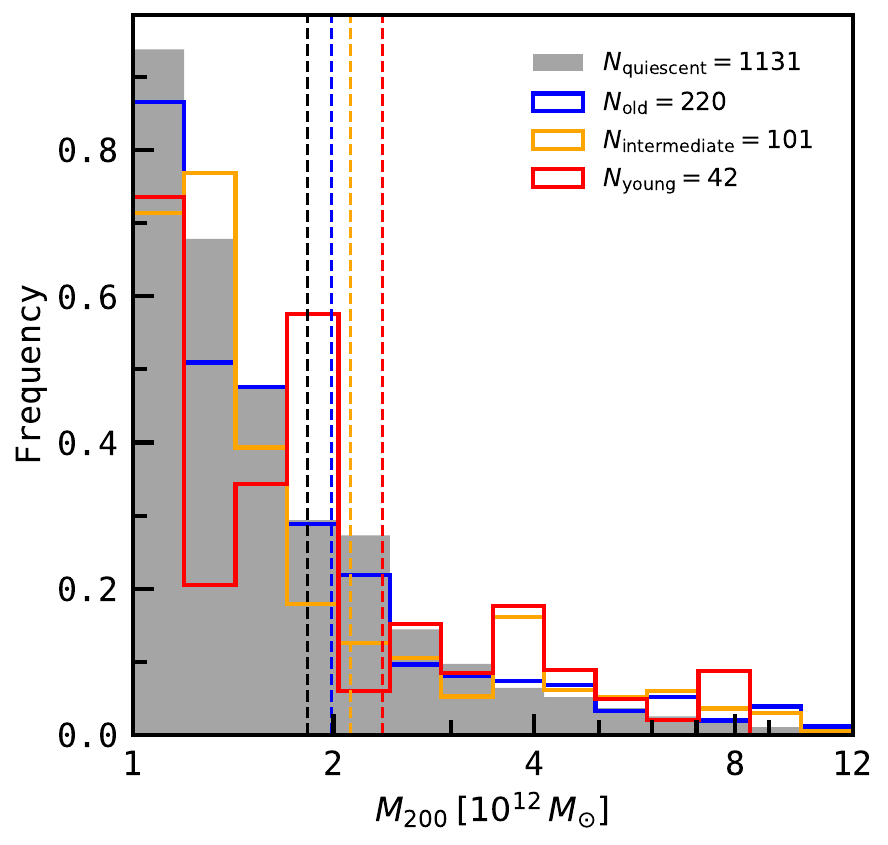}
    \hspace{5mm}
	\includegraphics[width=0.4\textwidth]{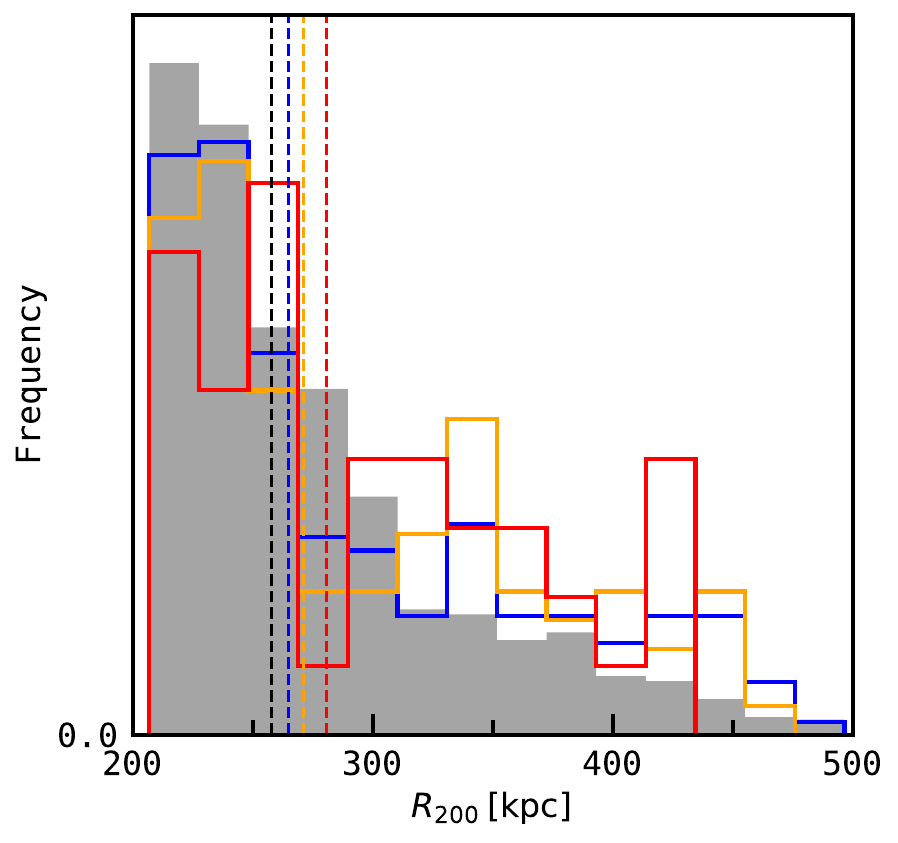}
    \caption{
        The virial mass and virial radius distribution of sampled host haloes, categorised by the lookback time to their last major merger -- dashed lines in the respective colour represent median values for each subsample.
    }
    \label{fig:aa_scale_byhistory}
\end{figure*}

In this Appendix, we present several additional scaling relations that are of importance to our conclusions made in the main paper.

In Fig.~\ref{fig:aa_extent_ca}, the flattening of satellite distributions appear to be enhanced for spatially extended systems when holding the satellite population constant. \emph{Full}-sample systems demonstrate a negative correlation between $c/a$ and $d_{\mathrm{rms}}/R_{200}$ (Kendall: $\tau=-0.34$, $p=5\times10^{-90}$), whereas no statistically significant correlation is found for systems in the \emph{virial} sample (Kendall: $\tau=0.01$, $p=0.56$). Flattening in simulated systems appear to be driven by distant satellites beyond their host's virial radius. We additionally confirm that only our specified satellite search radius has a major impact on such a correlation by plotting rolling averages for systems in the \emph{progenitor}, \emph{virial progenitor}, and \emph{low} samples (see Table~\ref{tab:s3_categories} for definitions). Our conclusions for \emph{full} and \emph{virial} systems also hold for their counterparts ranked by maximum progenitor mass. Reducing the number of satellites maintains the enhanced contribution of the outermost satellites to their distribution's flattening, while simultaneously having the expected effect of lowering $c/a$ in general.

In Fig.~\ref{fig:aa_scale_byhistory}, we show that the mass and extent of haloes are generally correlated with the recency of their last major merger. \emph{Young} haloes, which have experienced a merger in the last 2 Gyr, demonstrates a median $M_{200}$ around $30$ per cent larger than that of \emph{quiescent}-type haloes which have not experienced a major merger within the last 10 Gyr -- and a corresponding 23 kpc increase in $R_{200}$.

\section{Host galaxy morphology}
\label{sec:ab}

We now check whether a merger's impact on a satellite distribution is dependent on the visual morphology of the system's host galaxy at present time. We solely consider systems from TNG100 due to the availability of deep learning-based galaxy morphologies from \citet{Huertas-Company2019morph} -- specifically in the form of a variable $P_{\mathrm{late}}$, which represents the likelihood that a given galaxy is late-type. We classify our TNG100 hosts sampled in Section~\ref{sec:s2_sampling} into 832 early-type and 657 late-type galaxies with a threshold of $P_{\mathrm{late}}=0.5$, of which 184 and 176 respectively are \emph{merger-type} (i.e. have experienced a major merger in the last 10 Gyr).

The distribution of merger properties shown in Fig.~\ref{fig:s3_mergerproperties} are statistically indistinguishable between satellites of early and late-type hosts, with two-sample Kolmogorov-Smirnov $p$-values of 0.99 (lookback time), 0.54 (merger ratio), and 0.08 (merger duration). When comparing phase-space correlation metrics between systems around early and late-type hosts with respect to the recency of their last major merger (as in Fig.~\ref{fig:s3_byhistory}), $c/a$ and both kinematic metrics are statistically indistinguishable for all categories (\emph{quiescent}, \emph{old}, \emph{intermediate}, and \emph{young}) -- see Table~\ref{tab:aa_merger_ks}.

On the other hand, we recover a tendency for late-type central galaxies to host satellite distribution with lower plane heights and smaller radial extents. This trend is most significant for \emph{quiescent} systems (with shifts in mean $\Delta_{\mathrm{rms}}$ and $d_{\mathrm{rms}}/R_{200}$ of 15 kpc and 0.05 respectively), and grows less prominent for systems with recent mergers -- becoming statistically indistinguishable entirely (KS: $p > 0.05$) for \emph{young} systems (although this is also a consequence of a much smaller available population). This discrepancy in $\Delta_{\mathrm{rms}}$ is driven purely by late-type hosts tending towards smaller halo masses -- with a median $M_{200}$ of $2.23\times10^{12}\,M_{\odot}$ and $1.46\times10^{12}\,M_{\odot}$ for early- and late-type hosts respectively -- and thus $d_{\mathrm{rms}}/R_{200}$, while the scale-free flattening ($c/a$) is not significantly affected by host morphology.

In general, the disruptive impact of mergers on the phase-space correlation of the satellite systems involved appears to wash out any dependence on host morphology, while the overall flattening ($c/a$) and kinematic coherence ($\theta_{\mathrm{plane}}$, $\theta_{\mathrm{orbit}}$) follows the same distribution for both early and late-type hosts.

\begin{table}
	\centering
	\caption{Two-sample Kolmogorov-Smirnov tests comparing phase-space metrics of systems hosted by early and late-type hosts. Comparisons are performed for each of the four system categories defined in Table~\ref{tab:s3_categories}, and corresponding $p$-values (i.e. the likelihood that both metric distributions were drawn from the same parent sample) are shown -- those for which $p \leq 0.05$ (considered here to be statistically distinguishable) are marked in bold. The sample size of early-type and late-type hosts with respect to their merger history is shown below the corresponding category labels.}
	\vspace*{-1mm}
	\begin{tabular}{lllll}
		\hline
		Metric & Quiescent & Old & Intermediate & Young\\
        \hline
        $N_{\mathrm{early}}$ & 653 & 112 & 52 & 20\\
        $N_{\mathrm{late}}$ & 478 & 108 & 49 & 22\\
		\hline
		$c/a$ & 0.54 & 0.52 & 0.72 & 0.79\\
        $\Delta_{\mathrm{rms}}$ & $\mathbf{6.5\times10^{-12}}$ & $\mathbf{1.3\times10^{-8}}$ & $\mathbf{1.4\times10^{-6}}$ & 0.49\\
        $d_{\mathrm{rms}}/R_{200}$ & $\mathbf{1.5\times10^{-6}}$ & 0.25 & 0.53 & 0.31\\
        $\theta_{\mathrm{plane}}$ & 0.59 & 0.90 & 0.38 & 0.38\\
        $\theta_{\mathrm{orbit}}$ & 0.06 & 0.71 & 0.81 & 0.16\\
		\hline
	\end{tabular}
	\label{tab:aa_merger_ks}
\end{table}


\bsp	
\label{lastpage}
\end{document}